# Non-modal disturbances growth in a viscous mixing layer flow

**Helena Vitoshkin** and **Alexander Yu. Gelfgat**

*School of Mechanical Engineering, Faculty of Engineering, Tel-Aviv University, Ramat Aviv, Tel-Aviv 69978, Israel.*

**Abstract**

Non-modal transient growth of disturbances in an isothermal viscous mixing layer flow is studied for the Reynolds numbers varying from 100 up to 5000 at different streamwise and spanwise wavenumbers. It is found that the largest non-modal growth takes place at the wavenumbers for which the mixing layer flow is stable. In linearly unstable configurations the non-modal growth can only slightly exceed the exponential growth at short times. Contrarily to the fastest exponential growth, which is two-dimensional, the most profound non-modal growth is attained by oblique three–dimensional oblique waves propagating at an angle with respect to the base flow. By comparing results of several mathematical approaches, it is concluded that within the considered mixing layer model with the *tanh* base velocity profile, the non-modal optimal disturbances growth results from the discrete part of the spectrum only. Finally, full three-dimensional DNS with the optimally perturbed base flow confirms the presence of the structures determined by the transient growth analysis. The time evolution of optimal perturbations is presented and exhibit growth and decay of flow structures that sometimes become similar to those observed at late stages of time evolution of the Kelvin-Helmholtz billows. It is shown that non-modal optimal disturbances yield a strong mixing without a transition to turbulence.

# 1. Introduction

It is well known that in parallel shear flows there can be a transient amplification (or growth) of a disturbance energy even when all of the eigenvalues of the linearized problem indicate a perturbations decay. This phenomenon, which takes place at relatively short times, is widely recognized as "temporal" or "non-modal" growth. The growth is caused by a non-orthogonality of the flow eigenmodes and the result is independent of whether or not shear flow is unstable to exponential growth (see, e.g., Ellingsen & Palm, 1975, Landahl, 1980, Schmid & Henningson, 2001).

The non-modal disturbances growth is well studied for bounded shear flows, e.g., for plane Couette and Poiseuille flows (Farrel, 1987,1988; Buttler & Farrell, 1992; Reddy & Henningson, 1993; Schmid & Henningson, 2001). The semi-unbounded Blasius boundary layer flow is extensively investigated as well (see, e.g., Andersson *et al.*, 1999; Schmid & Henningson, 2001; Åkervik *et al.*, 2008). However, the problem of non-modal growth in fully unbounded flows, such as a mixing layer and a jet flow, is not completely understood. In this paper we consider the problem of transient non-modal growth of disturbances in a mixing layer flow with a hyperbolic tangent velocity profile. This flow is unstable within the inviscid model, and becomes linearly unstable at rather low Reynolds number if the viscous flow model is implied. The flow remains linearly stable for the streamwise wavenumber larger than unity (Drazin, 2001; Gelfgat & Kit, 2006). However, the instability development at early times still can be subject to a non-modal growth, so that the issue should be studied for the mixing layer flow as well.

The initial growth in the inviscid mixing layer flow was studied by Bun & Criminale (1994) and Criminale *et al.* (1995), who showed that a temporal perturbation growth in a mixing layer is possible. Later Le Dizes (2003) examined the non-modal growth of two-dimensional disturbances in inviscid and viscous mixing layer flows. It is shown in the present paper that the growth of three-dimensional perturbations is expected to be larger. Yecko *et al.* (2002) and Yecko & Zaleski (2005) studied the non-modal growth in a two-phase mixing layer for three-dimensional disturbances and found that the largest non-modal growth results from two-dimensional perturbations located in the spanwise plane and is uniform in the streamwise direction. Heifetz and Methven (2005) interpreted the optimal perturbation growth in an inviscid mixing layer in terms of the counter propagating Rossby waves. Recently, Bakas and Ioannou (2009) studied non-modal growth of two-dimensional disturbances in an inviscid mixing layer with a free surface. All these papers focused on the early transient disturbances evolution in the mixing layer flow. Surprisingly, the non-modal three-dimensional growth of a single-phase viscous mixing has not been addressed until now, except some preliminary results of Vitoshkin et al. (2012). Since it is common knowledge nowadays that the fastest disturbance growth in shear flows is defined by the optimal non-modal perturbation, we believe that the issue has to be studied also for the classical mixing layer flow model. A rather strong initial growth was observed in experiments of Gaster *et al.* (1985) and Kit *et al.* (2007)[1], which motivates our study additionally. In the course of this study, we did not discover any surprisingly large non-modal transient growth. However, we believe that the results reported below complement to the common understanding of the mixing layer flow properties and its behavior at early times. Besides that, we show that if perturbations wavelengths can be externally controlled as in, e.g., Gaster *et al.* (1985) and Gelfgat & Kit (2006), the non-modal growth can be used as a means of effective mixing with keeping the flow fully laminar.

In this paper, a transient non-modal growth of disturbances in a mixing layer flow is considered and analyzed numerically. This flow is known to be linearly unstable either if the inviscid flow is considered, or starting from rather low Reynolds numbers when the viscous flow model is implemented. The numerical code, based on the finite difference discretization of the Orr-Sommerfeld and Squire equations, is verified against well-known results on plane Poiseuille and Blasius boundary flows.

Orr-Sommerfeld and Squire equations are usually solved by spectral or pseudospectral methods (see, e.g., Schmid & Henningson (2001) and references therein). However, Gelfgat & Kit (2006) argued that steep changes in the $tanh$ velocity profile make it difficult to decompose the base flow as a series of convenient basis functions, such as, e.g., Chebyshev polynomials. The latter slows down the convergence and may lead to an undesirable Gibbs-like phenomena. Thus, in this study the Orr-Sommerfeld and Squire equations are discretized by the second-order finite difference method.

While performing computation for the mixing layer flow, we observed an unexpected loss of numerical accuracy. The codes validated against several simpler problems, e.g., plane Couette and Poiseuille flows, yielded unphysical results when the mixing layer $tanh$ velocity profile was substituted as the base flow. To overcome this difficulty for the Orr-Sommerfeld and Squire equations, we calculated their spectra with the quadruple precision (i.e., with 32 decimal places in the floating point numbers). For an additional verification, the numerical solution of time-dependent ODEs, as well as fully 3D Navier-Stokes equations are carried out starting from the base flow perturbed by the calculated optimal disturbance.

---

[1] Private communication with E. Kit

The convergence studies reported here show that an acceptable convergence can be reached by applying very fine and densely stretched grids with more than 1000 nodes in the cross-stream direction, which results in a large eigenvalue problem. Since the numerical model is bounded and its dimension is always finite, its spectrum is discrete. Analyzing the computed spectrum we observe a well-defined part that corresponds to the discrete spectrum of the initial unbounded problem. The amount of eigenmodes in this part remains constant independently on the grid refinement. Another part that should be attributed to the continuous spectrum of the unbounded problem converges extremely slowly and does not decay towards the boundaries of computational domain. To establish confidence in the obtained results on non-modal growth, we performed the computations using (i) the procedure offered by Reddy & Henningson (1993); (ii) the variational method offered by Butler & Farrel (1992); and (iii) the iterative forward/backward integration of governing/adjoint equations (Corbett & Bottaro, 2000). The approaches (i) and (ii) are applied for the discrete spectrum only, while approach (iii) includes the entire spectrum. Since all three approaches yield the same growth functions, we conclude that the continuous spectrum plays no role in the non-modal growth of the mixing layer flow. This conclusion is supported by calculations of the $\epsilon$-pseudospectrum (Trefethen & Embree, 2005; Mao & Sherwin, 2012). Furthermore, it was verified by monitoring of the energy growth calculated via the ODEs IVP problem, and the fully 3D time-dependent Navier-Stokes solution, both of which do not make any assumptions about the spectrum. We also suggest several additional arguments for exclusion of the reminiscence of continuous spectrum from the present non-modal analysis.

Following the time evolution of 2D and 3D optimal perturbation patterns, we observe that initially they are tilted against the shear slope and during the time evolution transform into a set of structures aligned along the shear. Non-modal analysis revealed that 3D perturbations are developing in different way and attain larger non-modal growth than correspondent 2D perturbation. The mechanistic interpretation for this phenomenon is given in Vitoshkin et al. (2012). Finally, we generate the initial data for three-dimensional direct numerical simulations using calculated three-dimensional optimal perturbations. Fully non-linear 3D computations allow us to confirm the previous findings, as well as to explore non-linear evolution of optimal disturbances into the viscous mixing layer flow. We show that initially small-amplitude optimal disturbance can grow so that non-linear terms become significant, which leads to formation of flow structures qualitatively different from the well-known Kelvin-Helmholtz billows at early stages of the instability onset. The optimal disturbances grow and decay in time yielding, in particular, a significant mixing inside the shear zone. It is quite an exceptional case of mixing since it is not followed by any transition to turbulence, which may be practically important.

Comparing the above flow structures with the experimental and numerical results on the developing mixing layer flows, we have found that similar flow patterns are observed at late stages of non-linear development of the Kelvin-Helmholtz instability. We argue that at long times after the linear instability onset, the effective width of the mixing layer grows so that the wavelength scaled by the width diminishes, while the corresponding wavenumber grows. As a result, the stable mixing layer configuration is created. This configuration is necessarily perturbed by the time-developing flow, which can trigger the non-modal growth resulting in similar flow structures.

In the following we give a brief formulation of the problem (Section 2) and describe the solution techniques applied and the test calculations made (Section 3). In Section 4 we discuss the effect of discrete and continuous spectra on the non-modal growth in the considered flow. Main results are presented in Section 5. We start from the growth functions and the optimal perturbation patterns yielded by the non-modal analysis. Then we study time evolution of the optimal disturbances within linear and non-linear, two- and three-dimensional models. Conclusions are summarized in Section 6.

## 2. Problem formulation

We consider an isothermal incompressible mixing layer flow produced by two fluid layers moving with opposite velocities $\pm U_{max}$ in the $x$-direction. Assuming that the mixing layer characteristic width is $\delta_v$, the hyperbolic tangent velocity profile $U(z) = U_{max} \tanh(z/\delta_v)$ is taken as a base flow. We are interested in temporal evolution of a small three-dimensional disturbance $\mathbf{v}=(u,v,w)^T$, which is governed by the non-dimensional momentum and continuity equations

$$\left(\frac{\partial}{\partial t} + U(z)\frac{\partial}{\partial x}\right)\mathbf{v} + \frac{dU}{dz}w\hat{\mathbf{e}}_x + (\mathbf{v}\cdot\boldsymbol{\nabla})\mathbf{v} = -\boldsymbol{\nabla}p + Re^{-1}\Delta\mathbf{v}, \qquad (1)$$

$$\boldsymbol{\nabla}\cdot\mathbf{v} = 0.$$

Here $\mathbf{v}=(u, v, w)$ is the velocity with components in the streamwise ($x$), spanwise ($y$) and vertical ($z$) directions; $p$ is the pressure; $\Delta$ denotes the vector Laplacian operator. The equations are rendered dimensionless using the scales $\delta_v$, $U_{max}$, $\delta_v/U_{max}$, and $\rho U_{max}^2$ for length, velocity, time and pressure, respectively. The Reynolds number is defined by $Re = U_{max}\delta_v/\nu$, where $\nu$ is the kinematic viscosity.

The flow is assumed to be periodic in the spanwise and streamwise directions, so that we consider the normal mode expansion and study solutions with fixed wavenumbers $\alpha$ and $\beta$ in the *x*- and *y*- directions. Since the temporal stability problem is considered, both wavenumbers are real. Looking for the infinitesimal perturbations of the base flow in the form $\{u(z,t), v(z,t), w(z,t), p(z,t)\} exp[i(\alpha x + \beta y)]$ and using standard derivation procedure we arrive to the set of Orr-Sommerfeld (OS) and Squire equations:

$$\Delta \frac{\partial w}{\partial t} = i\alpha \left(\frac{d^2 U}{dz^2} w - U\Delta w\right) + \frac{1}{Re}\Delta^2 w, \qquad (2)$$

$$\frac{\partial \eta}{\partial t} = -i\beta \frac{dU}{dz} w + \left(\frac{1}{Re}\Delta - i\alpha U\right)\eta. \qquad (3)$$

in which the vertical components of velocity *w* and vertical component of vorticity $\eta$,

$$\eta = \partial v/\partial x - \partial u/\partial y \qquad (4)$$

(see e.g., Schmid & Henningson, 2001). Here the Laplacian operator reduces to $\Delta = \frac{\partial^2}{\partial z^2} - (\alpha^2 + \beta^2)$. The problem is considered for *t*>0 and $-L \leq z \leq L$, where *L* must be large enough to ensure results independence on further increase of *L*. To make our analysis compatible with the previous numerical studies (e.g., Rogers & Moser, 1992; Kit *et al.*, 2010) we assume that all the perturbations vanish at $z = \pm L$.

In the following we study initial temporal growth of a perturbation in terms of kinetic energy norm, produced by the corresponding inner product (the star denotes the complex conjugate):

$$E(t) = \langle \mathbf{v}, \mathbf{v}\rangle = \int_V \mathbf{v}^* \cdot \mathbf{v}\, dV, \qquad \text{where } \langle \mathbf{u}, \mathbf{v}\rangle = \int_V \mathbf{u}^* \cdot \mathbf{v}\, dV \qquad (5)$$

We define the optimal disturbance as one yielding the maximum possible amplification of its initial energy norm. Following Farrell (1987, 1988) and Butler & Farrell (1992), the maximal amplification is defined as the maximal possible growth of the perturbation norm at a given time *t* and is considered for a single particular set of stability parameters (*α, β, Re*). The energy amplification, or growth function *G(t)*, is defined as:

$$G(t) = \max_{E(0) \neq 0} \frac{E(t)}{E(0)} \qquad (6)$$

Clearly, the above formulation remains meaningful only at relatively small times before the viscosity effects widen the flow profile. To estimate these meaningful times for different Reynolds numbers, we consider a simple model described in the Appendix A, where we show, e.g., that $t < 30$ remains meaningful for $Re = 1000$.

# 3. Solution technique and test calculations

Owing to the reasons described in the Introduction the equations (2) and (3) were discretized using the second order central finite difference schemes. After discretization, the governing equations are reduced to a system of linear ODEs governed by a matrix **L** assembled from all the discretized equations. The spectrum and the eigenvectors of **L** were computed using the QR algorithm. The transient growth is studied by three different numerical approaches: (i) using factorization of the Gram matrix (Reddy & Henningson, 1993 and Henningson & Schmid, 2001) and singular value decomposition (SVD); (ii) applying the calculus of variations (Butler & Farrell, 1992); (iii) by iterative forward/backward integration of the governing/adjoint equations (Corbett and Bottaro, 2000). All the three methods are implemented to cross-verify the results, as well as to support conclusions of Section 4.

For the code verification, we calculated the critical energetic Reynolds number and growth function for the plane Poiseuille flow and Blasius boundary layer profile (Table 1). The results are well compared with the published data of Reddy & Henningson (1993) and Schmid (2000). In both cases, using 600 nodes grid, we observed convergence up to the fourth decimal place at least, and even slightly improved the previous results.

Calculations for the mixing layer flow appear to be significantly more difficult. We observed, for example, that in spite of well-known stable numerical properties of the QR decomposition, calculations with the quadruple precision (i.e, 32 decimal places for floating point numbers) are needed to calculate the spectrum accurately. Note that taking the complex conjugate of the eigenvalue problem together with the transformation $z \to -z$, one can show that anti-symmetry of the base velocity profile implies appearance of complex eigenvalues in conjugated pairs (Appendix B). The corresponding eigenvectors *are not* complex conjugated, but are located in the opposite midplanes $z \geq 0$ or $z \leq 0$. Use of the double precision instead of the quadruple, one leads to spurious numerical errors, which can be seen, for example, as an appearance of non-conjugated pairs of complex eigenvalues.

The computational grid was divided into two parts. A half of the grid points were located inside the interval $-2 \leq z \leq 2$ and were stretched towards the centerline $z=0$. The stretching function used is $\tanh(sy)/\tanh(s)$. The fastest convergence was observed for $s=3$. Remaining parts of the grid above and below the interval $-2 \leq z \leq 2$ were uniform. This grid arrangement yields a strong stretching near the mixing zone, where the linearly most unstable eigenvectors are located (see, e.g., Gelfgat & Kit, 2006). Outside the mixing zone the discrete spectrum eigenvectors decay, so that there an unnecessary stretching is removed. It is emphasized that only this grid arrangement allowed us to obtain grid-independent results with the use of 1000-2000 grid points. Use of continuously stretched or uniform grids with the same amount of grid nodes exhibited an unacceptable grid-dependence (see Table 3).

Table 1. Convergence of critical energetic Reynolds numbers $Re_{crE}$, and the growth function $G(t)$. Comparison with results of Reddy & Henningson (1993) and Schmid (2000).

| Order of grid, ($N$) | Poiseuille flow | | | | Blasius boundary layer | |
|---|---|---|---|---|---|---|
| | $Re_{crE}$ $\alpha=0, \beta=1.9$ | $Re_{crE}$ $\alpha=3.2, \beta=3$ | $G_{max}$ $\alpha=1, \beta=0$ $Re=3000$ | $G_{max}$ $\alpha=0.5, \beta=2.5$ $Re=1000$ | $G_{max}$ $\alpha=0.1, \beta=0.26$ $Re=1000$ | $G_{max}$ $\alpha=0.2, \beta=0.47$ $Re=1000$ |
| 200 | 49.94 | 88.83 | 21.87 | 198.23 | 221.96 | 394.82 |
| 300 | 49.89 | 88.64 | 21.47 | 198.01 | 219.89 | 397.06 |
| 400 | 49.87 | 88.57 | 20.95 | 197.57 | 213.06 | 398.11 |
| 500 | 49.85 | 88.51 | 20.29 | 197.55 | 213.64 | 398.12 |
| 600 | 49.85 | 88.51 | 20.28 | 197.54 | 213.65 | 398.13 |
| 700 | 49.85 | 88.51 | 20.28 | 197.54 | 213.65 | 396.13 |
| Reddy & Henningson (1993) | 49.7 | 87.6 | 20.37 | 196 | | |
| Schmid | | | | | ≈200 | ≈400 |

The computational domain for the mixing layer flow is defined as an interval of width $2L$. According to recent results of Healey (2009) an insufficiently large value of $L$ can significantly alter flow stability properties. A series of test calculations for $L$ varying between 5 and 100 was carried out together with the necessary convergence study. Dependence of the leading eigenvalue and growth function value on the size of computational domain $L$ is presented in Table 2. Based on several similar calculations for different values of $\alpha$ and $\beta$, we concluded that the flow linear stability properties can be described correctly starting from $L=20$. This width of the computational domain corresponds also to the height of the experimental channel of Kit *et al.* (2007), and has been chosen for further computations.

Table 3 shows an example of convergence of four leading eigenvalues belonging to the discrete part of the spectrum. It is seen that use of 1000 grid points yields four converged decimal digits for the first mode, however, the convergence slows down for the next modes. This shows that the linear stability analysis (Gelfgat & Kit, 2004) is less computationally demanding than the non-modal growth study, for which several leading eigenmodes must be calculated within a good accuracy.

Table 2. Results for varying length of the computational domain $L$. $Re=1000$, calculation with 1000 stretched grid points.

| | 1st mode, $\lambda_i=0$, $\lambda_{real}$, | | Growth function, $G(t_{max})/t_{max}$ | |
|---|---|---|---|---|
| $L$ | $\alpha=1, \beta=0$ | $\alpha=0.7, \beta=1$ | $\alpha=1, \beta=0$ | $\alpha=0.7, \beta=1$ |
| 5 | -0.0427 | -0.1102 | 98.15 / 15.5 | 782.94 / 24.7 |
| 10 | -0.0425 | -0.1102 | 98.41 / 15.7 | 784.52 / 24.7 |
| 15 | -0.0425 | -0.1101 | 98.41 / 15.7 | 784.18 / 24.9 |
| 20 | -0.0424 | -0.1101 | 98.41 / 15.7 | 784.04 / 24.9 |
| 30 | -0.0424 | -0.1101 | 98.41 / 15.7 | 784.04 / 24.9 |
| 50 | -0.0424 | -0.1101 | 98.41 / 15.7 | 784.04 / 24.9 |
| 100 | -0.0424 | -0.1101 | 98.41 / 15.7 | 784.04 / 24.9 |

Table 3. Convergence of four least stable eigenvalues belonging to discrete spectrum for $\alpha=0.7$, $\beta=1$, $Re=1000$ (*tanh*-stretching divided mesh).

| $N$ | 1st mode | 2nd mode | 3rd mode | | 4th mode | | Growth function, |
|---|---|---|---|---|---|---|---|
| | $\lambda_r$, $\lambda_i=0$ | $\lambda_r$ | $\lambda_r$ | $\lambda_i$ | $\lambda_r$ | $\lambda_i$ | $G/t_{max}$ |
| 500 | -0.1116 | -0.2888 | -0.3022 | -0.1986 | -0.3727 | 0.3068 | 786.43 / 25.0 |
| 600 | -0.1102 | -0.3138 | -0.3231 | -0.1952 | -0.3452 | 0.3032 | 785.84 / 24.9 |
| 700 | -0.1102 | -0.3203 | -0.3287 | -0.1944 | -0.3504 | 0.3025 | 786.96 / 24.9 |
| 800 | -0.1102 | -0.3260 | -0.3339 | -0.1922 | -0.3551 | 0.3013 | 785.14 / 24.9 |
| 900 | -0.1102 | -0.3310 | -0.3371 | -0.1921 | -0.3589 | 0.3001 | 784.55 / 24.9 |
| 1000 | -0.1102 | -0.3385 | -0.3448 | -0.1905 | -0.3657 | 0.2987 | 784.04 / 24.9 |
| 1100 | -0.1102 | -0.3385 | -0.3445 | -0.1902 | -0.3653 | 0.2986 | 784.04 / 24.9 |
| 1200 | -0.1102 | -0.3384 | -0.3444 | -0.1902 | -0.3652 | 0.2986 | 784.04 / 24.9 |
| 1300 | -0.1102 | -0.3384 | -0.3443 | -0.1901 | -0.3652 | 0.2985 | 784.04 / 24.9 |
| 1400 | -0.1102 | -0.3383 | -0.3441 | -0.1901 | -0.3650 | 0.2985 | 784.04 / 24.9 |
| 1500 | -0.1102 | -0.3383 | -0.3441 | -0.1901 | -0.3650 | 0.2985 | 784.04 / 24.9 |

Another way to verify the calculated growth function is to calculate the solution to (2), (3) by integrating the ODEs with the optimal vector as the initial condition. In this case, the norm of the time-dependent solution at time *t* must be equal to the calculated growth function $G(t)$. In the following, the solution of the initial value problem is used for verification of the results, as well as to follow the time evolution of optimal vectors. For additional verification, we consider a fully non-linear time-dependent problem taking the optimal vector as an initial condition. The numerical technique used for solution of the 3D problem is described in Vitoshkin & Gelfgat (2012). Comparison of the kinetic energy evolution with the growth function calculated via the three independent approaches is shown and discussed below.

## 4. Spectrum of a linearized problem

As any flow in an unbounded domain, the mixing layer flow has two parts of the spectrum: a finite number of discrete eigenmodes and an infinite number of eigenmodes belonging to the continuous spectrum. Clearly, a numerical method, based on a discrete model defined for a bounded domain, cannot reproduce accurately the continuous modes. Grosh & Salven (1978) argued that continuous modes of the OS equation are either oscillatory or decaying functions located in free stream regions, where $U$=const and $U'$=0, and are zeroes in the regions where $U'\neq 0$. In our numerical results we observe similar modes that slowly decay or oscillate towards the ends of computational interval [-$L$,$L$]. Their amount grows with mesh refinement, however they are not exact zeroes in the shear zone (Figs. 1 and 2). The corresponding Gram matrix contains non-diagonal elements close to unity, which means that some modes are almost parallel with respect to the inner product (5). The latter can be expected for modes corresponding to the continuous spectrum of the unbounded problem. Furthermore, taking into account these almost parallel modes for computation of the growth function via the Gram matrix decomposition (Reddy & Henningson, 1993), results in a very large growth function reaching the values of the order of $10^{20}$, with the corresponding optimal vector located inside the uniform flow. Apparently, such a result is considered as unphysical and incorrect. Applying the calculus of variations, which is also based on the linearized problem spectrum, we arrive to a similar unphysical result. At this point we assume that the observed almost parallel modes are a non-accurate replication of the continuous spectrum. Furthermore, we argue that, (i) the procedures described in Sections 2.3.1 and 2.3.4 are applicable only to a finite number of eigenmodes; and (ii) the continuous modes do not decay far from the area of non-zero shear, so that the integral (2.2.7) does not necessarily converge for $L \rightarrow \pm\infty$, thus making all the procedures based on the chosen inner product meaningless, even if one assumes that discrete approximation of the continuous spectrum is sufficiently accurate. Therefore, the effect of the continuous spectrum on non-modal growth must be studied separately.

To account correctly for the problematic "continuous" eigenmodes, we apply the third approach, which is based on the forward/backward time integration of the governing/adjoint equations, and therefore necessarily takes into account the entire spectrum (Corbett & Bottaro, 2000). It can be seen (Table 4) that all the three approaches cross-verify each other and exhibit close results when applied to the Poiseuille flow, which is bounded and therefore has only discrete spectrum. At the same time, when using the first two methods for the whole calculated spectrum of the mixing layer flow, we obtain a very large unphysical non-modal growth of the order of $10^{15}$–$10^{20}$. Note, that the above observation disappears for the Blasius boundary layer flow, for which we leave all modes without separating them into discrete and continuous parts.

To apply the Gram matrix factorization / SVD approach, we extract the eigenvectors localized in the shear zone as is illustrated in the following example. Consider a certain set of parameters $Re$=1000, $\alpha$=0.7, $\beta$=1, for which we have non-modal growth. The calculated spectrum and examples of the eigenvector profiles are shown in Figs. 1 and 2. Three branches corresponding to the continuous spectrum are given by $Imag(\lambda) = 0$ and $Imag(\lambda) \approx \pm\alpha$ (Grosh & Salven, 1978). As expected, these sets of eigenvalues do not converge and their number increases with the grid refinement, which is indicative of their "continuous" origin. Conversely, the discrete eigenmodes can be recognized, primarily, by their fast convergence. Also, the amount of these eigenvalues remains constant with the grid refinement, indicating additionally on their "discrete" origin.

The discrete or continuous character of an eigenmode can be identified also by its eigenvector profile: the discrete eigenvectors are localized in the neighborhood of the mixing zone, while the continuous ones retain non-zero amplitudes far from the mixing area. Several examples are shown in Fig. 2. The eigenvalues corresponding to the plotted eigenvectors are numbered from 1 to 12 and are shown in the lower insert of Fig. 1. It is clearly seen that some eigenvectors do not decay far from the mixing zone, which is located in the interval $-3 \lesssim z \lesssim 3$ (Fig. 2b). We attribute these eigenmodes to the continuous spectrum and, following the above arguments, exclude them from further consideration. The modes decaying at large values of $|z|$, like those shown in Fig .2a, are attributed to the discrete spectrum and are included in the further analysis. For the following computations we exclude all the eigenmodes that are large for $|z| > 10$. We observe that the growth functions, as well as the number of extracted eigenvectors, do not change when boundaries of this interval vary from ±8 to ±15, which shows that the procedure is consistent. Remarkably, the number of extracted eigenvectors does not change also with the grid refinement (Table 4). This constant number of the eigenvectors localized near the shear zone can be a property of the discrete spectrum, however we have no clear criterion to define to which part of the spectrum an eigenvector belongs.

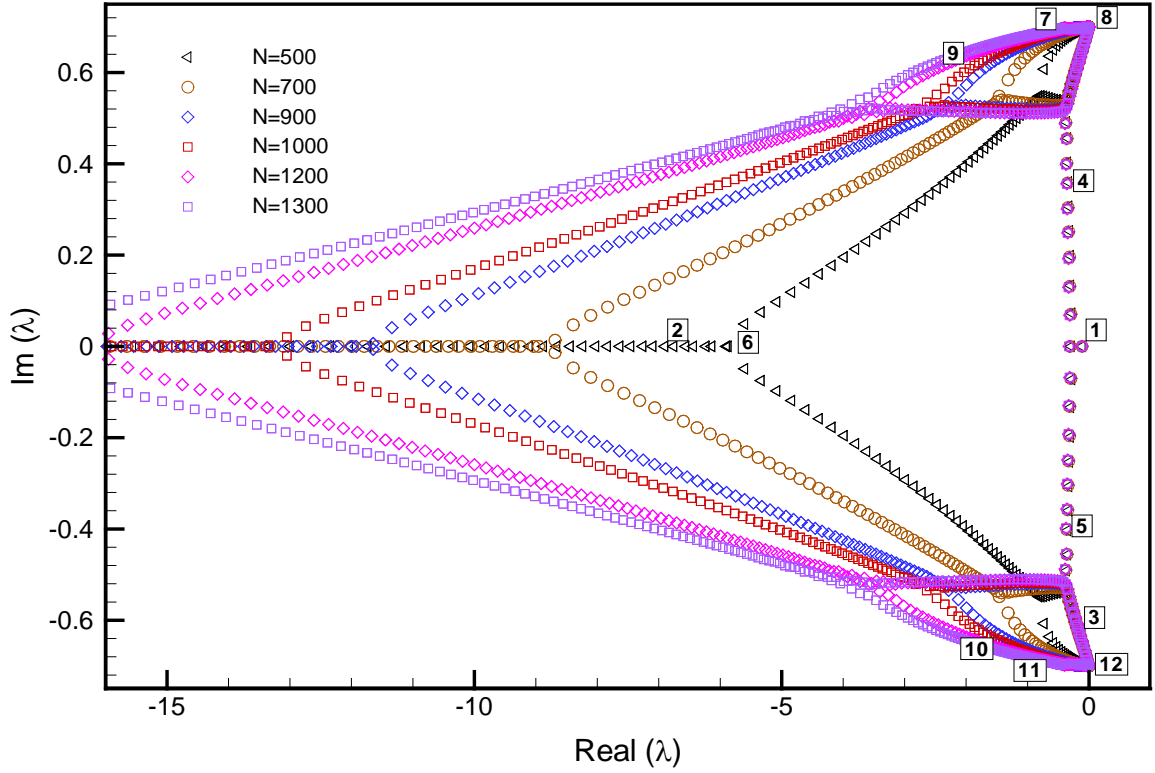

Fig. 1. Spectrum of the mixing layer flow at *Re*=1000, *α*=0.7, *β*=1 calculated for different numbers of grid points. Labeled points correspond to the eigenmodes shown in Fig. 2.

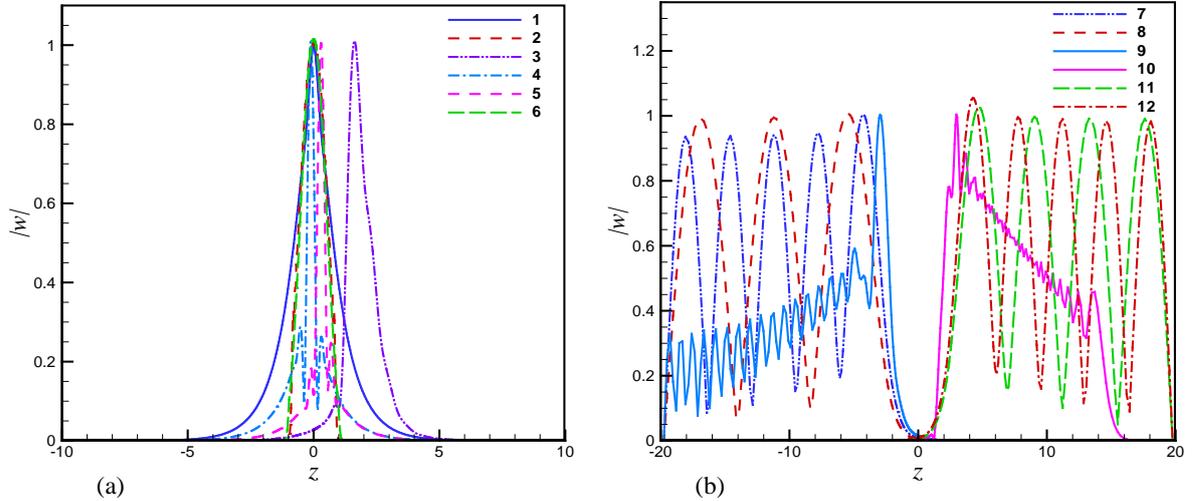

Fig. 2. Patterns of eigenvectors belonging to (a) discrete and (b) continuous parts of the spectrum. *Re*=1000, *α*=0.7, *β*=1. Number of line corresponds to the number of eigenvalues depicted in the lower insert of Fig. 1.

Following Mao & Sherwin (2011), we performed also the pseudospectrum analysis of the calculated spectrum. Figure 3a illustrates a calculated spectrum with the $\epsilon$-pseudospectrum computed as the minimal singular value of the matrix $(J - \lambda I)$, where $J$ is the Jacobian matrix of the discretized equations system and $\lambda$ is the current eigenvalue. We observe that the eigenmodes whose $\epsilon$-pseudospectrum is relatively large, $\epsilon > 10^{-4}$, do not decay in the free-stream region. The corresponding eigenvalues are characterized by $Imag(\lambda) \approx \pm\alpha$, which is also indicative of their "continuous" origin, therefore, these mode are excluded from calculations. At the same time, the eigenmodes whose $\epsilon$-pseudospectrum is bounded to $\epsilon < 10^{-5}$ coincide with the modes extracted according to the above arguments. An example of the growth functions calculated for only those modes whose pseudospectrum is bounded by either $10^{-5}, 10^{-6}, 10^{-7}$, or extracted as described above that corresponds to $\epsilon < 10^{-4}$, is presented in Figure 3b. It is seen that modes corresponding to $10^{-4} < \epsilon < 10^{-5}$ do not contribute to the optimal growth, while the modes whose pseudospectrum $\epsilon < 10^{-6}$ do influence it. Therefore, we can conclude that the eigenmodes corresponding to $\epsilon < 10^{-6}$ are those to be accounted for. It is emphasized, however, that we still have no clear criteria to separate continuous and discrete parts of the spectrum.

For an additional verification of our conclusion, we used the calculated optimal vectors as initial conditions for the ODE system (eqs. (3) and (4)), as well as fully 3D equations (1), and integrated them in time, monitoring the kinetic energy norm of the solution. We observed that at a chosen target time the solution norm reaches the calculated value of the growth function. Clearly, if parts of to-be-continuous modes were perturbed inside the mixing zone and were contributing into non-modal growth it would be impossible to obtain such a good agreement. In case a significant mode was mistakenly excluded, the real non-modal growth would be larger than the growth function calculated here. An example is shown in Figure 3c, where we compare the growth function calculated on the basis of the extracted discrete spectrum with the kinetic energy norm evolution yielded by time integration of the initial values ODEs and fully non-linear 3D problems. The equality of the maximal values of the norm and the growth function makes us confident in the results obtained, including the exclusion of the continuous spectrum.

The far right column of Table 3.3.1 shows results of the Gram matrix factorization / SVD approach of Reddy & Henningson (1993) and of the calculus of the variations method of Buttler & Farrell (1992), both applied to the extracted eigenmodes only. It is clearly seen that these results are identical and are very close to those obtained by the iterative time integration based method, which accounts for the whole spectrum.

We also verified behavior of the forward/backward time integration based method starting the iterations from two different initial vectors. The initial profiles were chosen as a wide parabola spreading into the uniform flow and the Gaussian function located inside the mixing zone. In both cases the same optimal vector was obtained after 5-6 iterations. We examined that this observation remains valid for different values of $Re$, $\alpha$ and $\beta$, and concluded that even when the free stream area is artificially perturbed, the optimal vector remains located within the shear zone. Thus, we can restrict the non-modal analysis to only those eigenvectors that vanish outside the shear zone. It is emphasized that having the spectrum computed, the Gram matrix factorization / SVD approach consumes significantly less CPU time than the one based on the forward/backward time integration, or than computation of an inverse energetic matrix needed for the variational method. This is an advantage, for example, when optimal growth at different target times is studied.

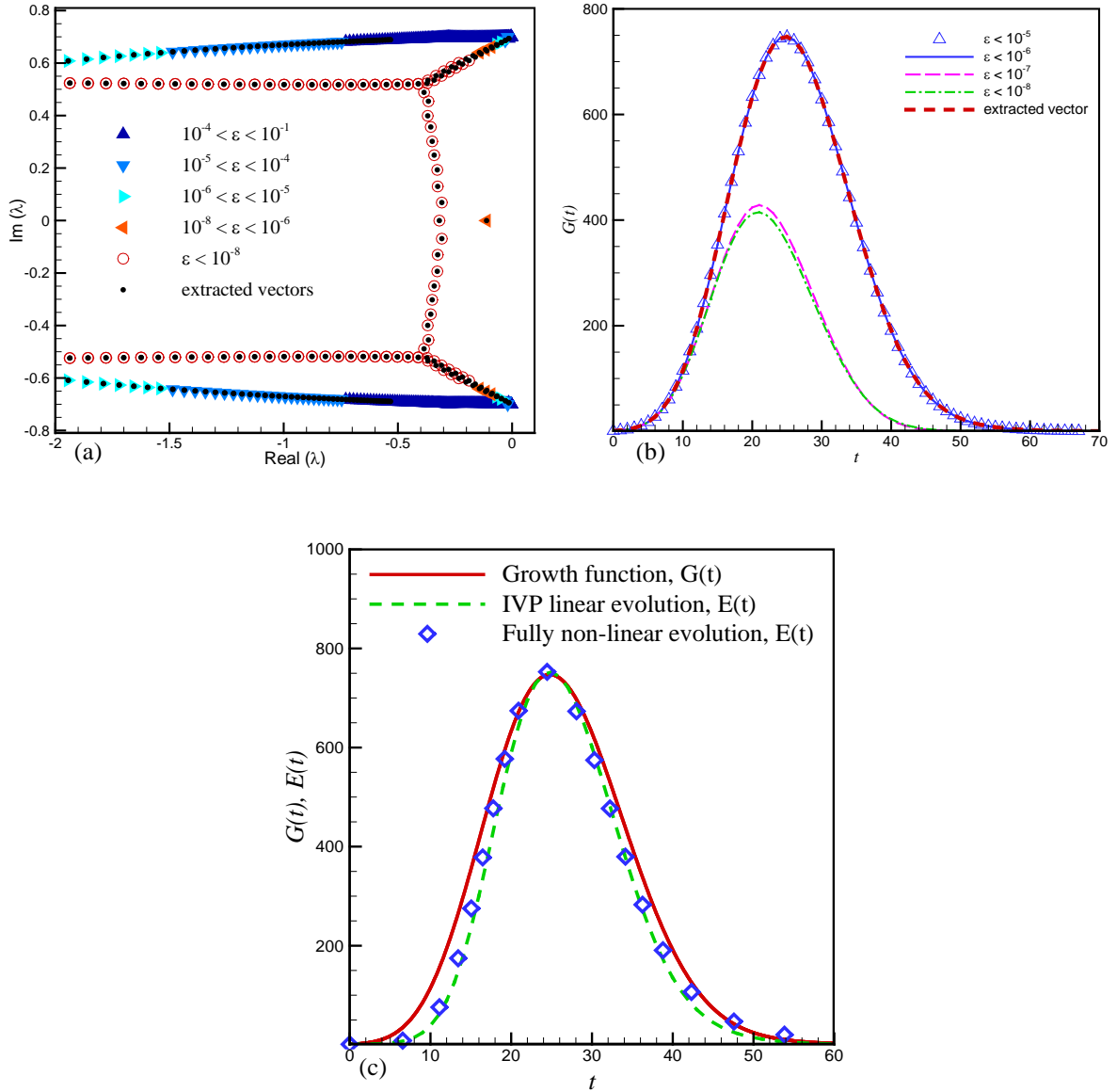

Fig. 3. (a) Spectrum and pseudospectrum of the mixing layer flow at $\alpha = 0.7, \beta = 1, Re = 1000$. (b) Growth functions calculated for reduced parts of spectrum corresponding to different values of $\epsilon$−pseudospectrum: (c) Comparison of the growth function $G_E(t)$ (solid line) with evolution of the kinetic energy of the optimal initial vector obtained as a solution of the ODEs IVP (dashed line) and as a solution of fully non-linear 3D problem (symbols).

Table 4. Growth function, $G(t=5)$ calculated by different methods for Poiseuille flow, boundary layer Blasius profile, and mixing layer flow.

| Number of grid points | Poiseuille flow $\alpha=1$, $\beta=0$, $Re=3000$, (only discrete) | Blasius boundary layer $\alpha=0.125$, $\beta=0.3$, $Re=800$, (discrete & continuous) | Mixing layer $\alpha=1$, $\beta=0$, $Re=1000$, (extracted vectors) |
|---|---|---|---|
| *using factorization of the Gram matrix – SVD* | | | |
| 500 | 5.44 | 1.77 | 14.66 (318 vectors) |
| 1000 | 5.45 | 1.77 | 14.66 (318 vectors) |
| 1500 | 5.45 | 1.77 | 14.66 (318 vectors) |
| *applying the calculus of variations* | | | |
| 500 | 5.44 | 1.78 | 14.66 |
| 1000 | 5.45 | 1.77 | 14.66 |
| 1500 | 5.45 | 1.77 | 14.66 |
| *by iterative forward/backward integration of the governing/adjoint equations (the whole spectrum)* | | | |
| 500 | 5.44 | 1.76 | 14.66 |
| 1000 | 5.45 | 1.77 | 14.66 |
| 1500 | 5.45 | 1.77 | 14.66 |

# 5. Non-modal growth in the isothermal mixing layer flow

A possibility of non-modal growth, even at very small Reynolds numbers, becomes obvious after comparison of the energetic and linear critical Reynolds numbers (Joseph, 1976) calculated for the mixing layer flow and reported in Appendix C. Note that critical energetic Reynolds numbers relate to the kinetic energy growth at the initial time when the flow is not affected yet by the viscosity effects. In the following we study the non-modal growth varying the Reynolds number together with the streamwise and spanwise wavenumbers.

*5.1. Growth function.*

The study of non-modal growth, was started for two-dimensional disturbances ($\beta=0$), which, due to the Squire transformation, are most linearly unstable. Growth functions were calculated for $Re=100$, 1000, and 5000. Several examples are shown in Fig 4. At large times, $t>10$, in all the cases considered, the exponential growth of linearly unstable modes prevails the non-modal growth. At short times, $t<10$, the non-modal growth of linearly stable modes with the streamwise wavenumber $\alpha \gtrsim 0.9$ can slightly exceed the exponential growth of linearly unstable modes. As is shown below, this faster non-modal growth can lead to noticeable non-linear effects. An interesting observation is that among all modes exhibiting non-modal growth, the maximal one is attained by modes whose streamwise wavenumber $\alpha$ lays between values 0.9 and 1, i.e. the values that correspond to linearly stable modes in viscous ($0.9<\alpha<1$) and neutral modes in inviscid ($\alpha=1$) mixing layers (Gelfgat & Kit, 2006). It should be emphasized that non-linear numerical modeling, as well as experimental studies, are usually done at the values of $\alpha$ corresponding to linear instability, preferably with the largest time increment. The present results show that non-linear evolution of stable and close to being neutral modes is also worth exploring (see Section 5.3).

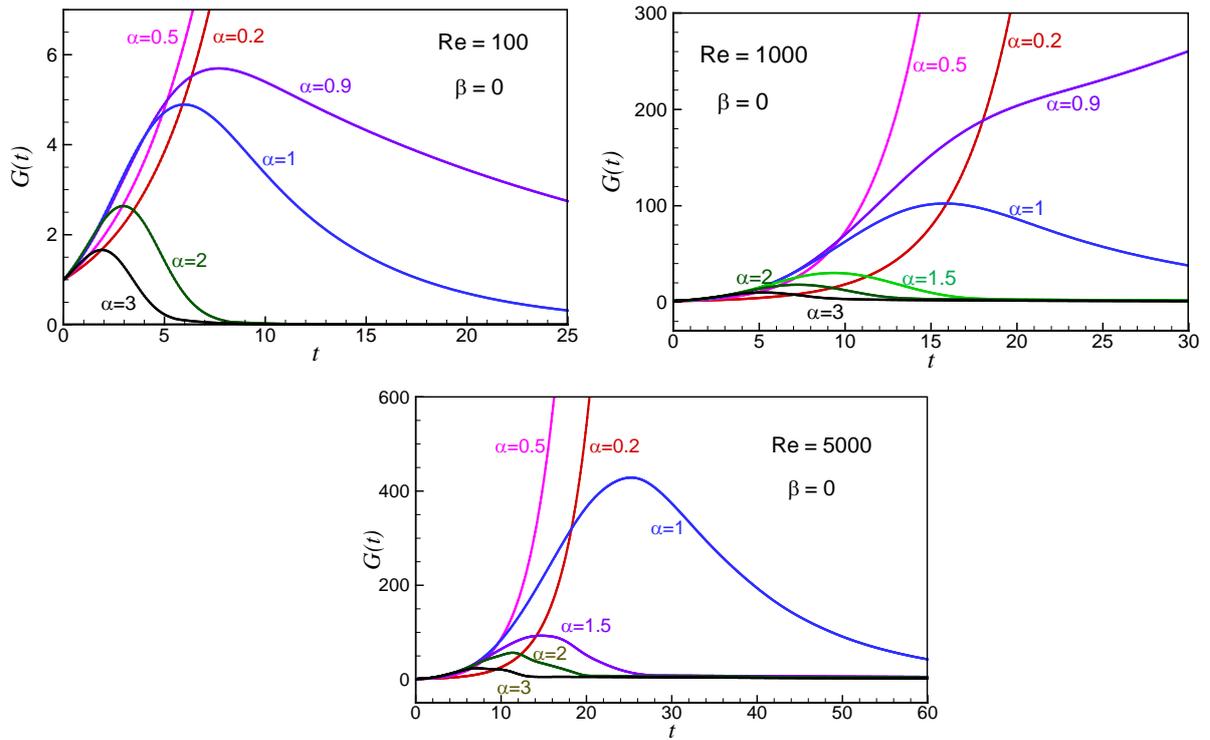

Fig. 4. Growth functions of two-dimensional disturbances, $\beta=0$.

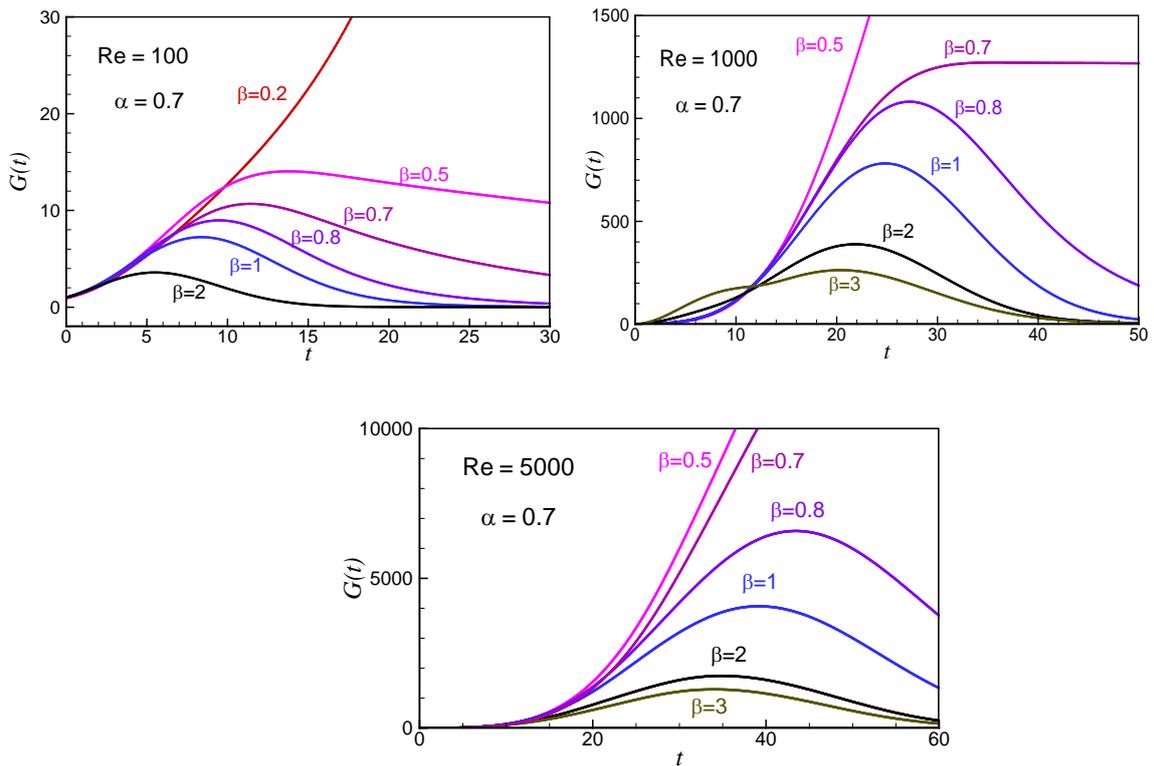

Fig. 5. Growth functions of three-dimensional disturbances for fixed $\alpha=0.7$.

Growth functions of three-dimensional disturbance modes at fixed value of the streamwise wavenumber $\alpha=0.7$ and varying spanwise wavenumber $\beta$ are shown in Fig. 5. Note that the two-dimensional mode corresponding to $\beta=0$ is linearly unstable also in this case. As before, the calculations were performed for $Re$=100, 1000 and 5000. Also in these cases we observe that for $Re\geq100$ the non-modal growth can slightly exceed the linear one at short times, while at longer times, the linear exponential growth always prevails. We observe also that the three-dimensional perturbations become stable when the spanwise wavenumber $\beta$ exceeds a certain value close to 0.5. The most noticeable non-modal growth can be attributed to the values of $\beta$ corresponding to the linearly neutral configuration (Fig. 5), when there exists at least one eigenmode slowly decaying in time. This observation sustains when we consider other fixed values of $\alpha$ and different $\beta$ as in Fig. 5. Table 5 summarizes the maximal values of the growth function $G_{max}$ and the times at which the maximum is attained $t_{max}$ over all wave numbers for 2D and 3D cases. The values of the $G_{max}$ at different $\alpha$, $\beta$, and $Re$ are plotted in Fig. 6.

Table 5. Maximal values $G_{max}$ and $t_{max}$ and corresponding wavernumbers.

| $Re$=100 | | | | $Re$=500 | | | | $Re$=1000 | | | | $Re$=5000 | | | |
|---|---|---|---|---|---|---|---|---|---|---|---|---|---|---|---|
| $\alpha$ | $\beta$ | $G_{max}$ | $t_{max}$ | $\alpha$ | $\beta$ | $G_{max}$ | $t_{max}$ | $\alpha$ | $\beta$ | $G_{max}$ | $t_{max}$ | $\alpha$ | $\beta$ | $G_{max}$ | $t_{max}$ |
| 0.9 | 0 | 5.7 | 7.7 | 0.9 | 0 | 25.3 | 11.7 | 1.0 | 0 | 102.1 | 15.5 | 1.0 | 0 | 428.4 | 24.9 |
| 0.5 | 0.7 | 14.6 | 13.0 | 0.5 | 0.8 | 3.8e3 | 27.8 | 0.5 | 0.8 | 1.6e4 | 26.8 | 0.5 | 0.8 | 9.9e4 | 42.9 |

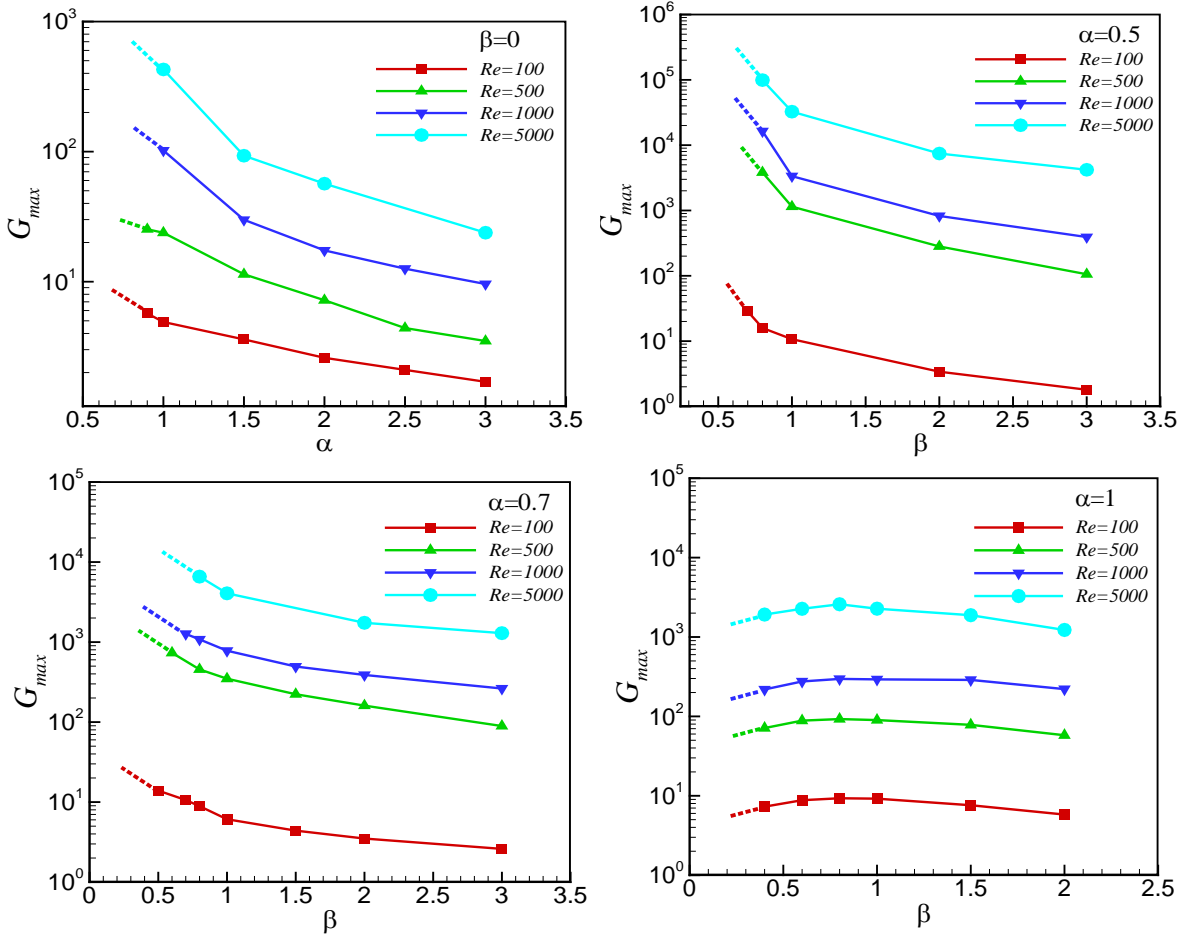

Fig. 6. Values $G_{max}$ and $t_{max}$ for different $\alpha$, $\beta$, and $Re$ (dashed line depicts linearly unstable flow).

We observe here that the largest non-modal growth is attained by oblique waves propagating at ($\alpha$=0.5, $\beta$=0.8), with respect to the base flow. The well-known non-modal growth studies for Couette flow (Butler & Farrell, 1992), Poiseuille flow (Reddy & Henningson, 1993) and Blasius flow (Schmid, 2000) also show that 3D perturbations exhibit the largest non-modal growth. This seems to be a common property of plane-parallel shear flows, which is discussed in detail by Vitoshkin et al. (2012).

### 5.2. Optimal vector

Figure 6 illustrates amplitudes and phases of the optimal vector for $\alpha$=0.7, $\beta$=0.8, $Re$=1000, the parameters characteristic for 3D transient growth. At these parameters the flow is linearly stable, while non-modal growth functions attain the maximal values close to the largest value over all possible spanwise wavenumbers. This choice allows us also to follow the time evolution of optimal vectors that will not be altered by an exponentially growing perturbation. The optimal vectors are calculated for the target time $t_{max}$=27.2, at which corresponding growth functions attain their maximal values. For an additional verification of our results, we used the optimal vector as an initial condition for (2), (3) and ensured that the time integration arrives to the final vector shown in Fig. 7, as predicted by the first left singular vector of the corresponding SVD. The growth of amplitude yielded by the IVP solution also coincides with the predicted growth function value.

Comparing profiles of the optimal vectors with those of the leading eigenvectors (see, e.g., Fig. 3 in Gelfgat & Kit, 2006) we observe that the optimal disturbance profiles are narrower and steeper. Contrary to the eigenvectors, the optimal disturbances amplitudes are symmetric with respect to the mixing layer midplane.

To follow the optimal disturbances time evolution, we plot their spatial patterns developing in time in the framework of the linearized equations (3) and (4). We start exploring the patterns evolution from the two-dimensional case, $\beta$=0, and focus on the case $\alpha$=1.5, for which no linearly growing eigenmodes exist. Figure 8a illustrates time evolution of the spanwise vorticity component $\eta_y$. Note that the base velocity $U(z)$ is positive in the upper part of the frames and is negative in the lower part, so that the shear slope is directed from the lower left to the upper right corner. Note the striking similarity between the optimal perturbation of the mixing layer flow and those found by Farrell (1988) and Buttler & Farrell (1992) for Couette and Poiseuille flows. The optimal initial disturbance appears as patterns tilted against the mean flow shear. The patterns are rotationally symmetric with respect to their centers located at the midplane. Developing in time, the patterns retain the rotational symmetry and rotate, becoming aligned along the shear slope. The maximum of kinetic energy corresponds to the vertical alignment of the patterns. At later times, the patterns tilt along the shear and decay. This evolution of the optimal disturbance corresponds to the well-known Orr mechanism (Orr, 1907).

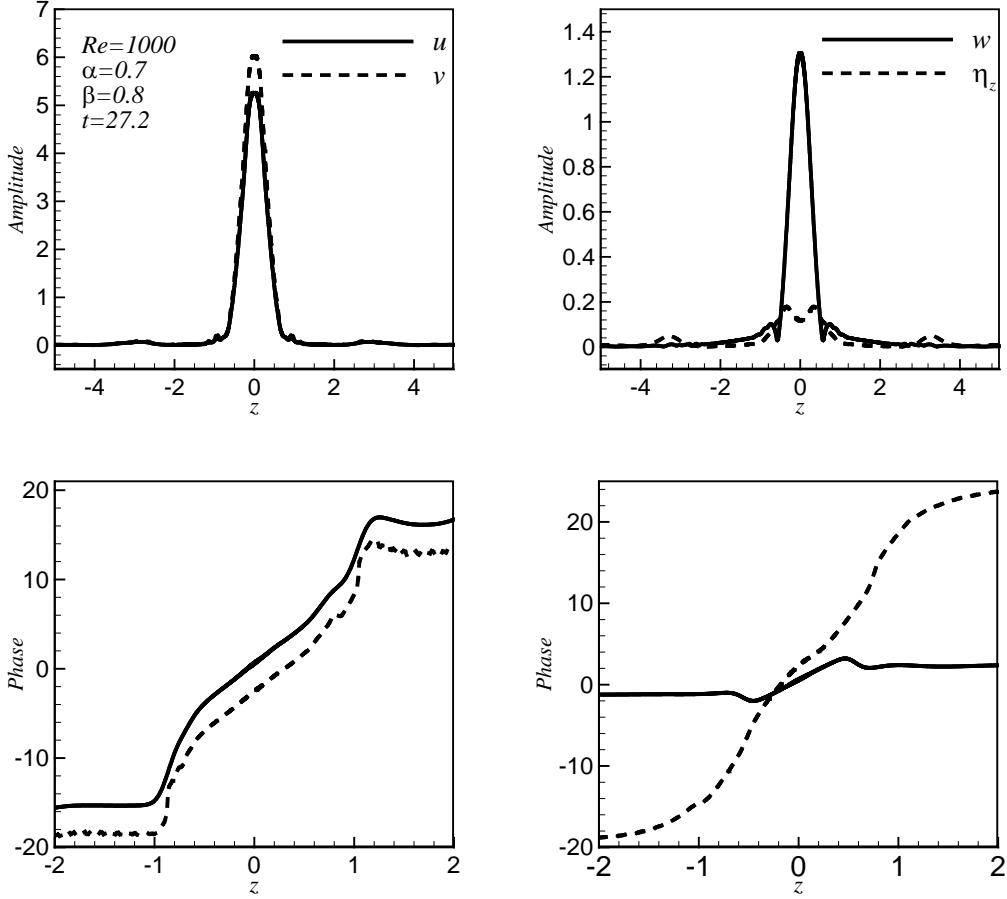

Fig. 7. Amplitudes and phases of the optimal disturbance vector for $Re=1000$, $\alpha=0.7$, and $\beta=0.8$ for times yielding the maximal values of growth functions ($t_{max}=27.2$).

Figure 8b shows time evolution of the spanwise vorticity component $\eta_y$ of the 3D optimal disturbance for the case $\alpha=0.7$, $\beta=0.8$, $Re=1000$. The disturbance pattern consists of a pair of rolls per one spatial period with their axes parallel to the vector $(2\pi/\alpha, 2\pi/\beta, 0)$. At the initial time, similarly to the 2D case, the rolls are tilted against the shear slope. During the time evolution the rolls grow and turn around their axes until reaching the position similar to the vertical alignment of the 2D rolls (Fig. 8a). After that the rolls are turning in the base flow direction, their kinetic energy continues to grow up to the maximum value, which is unlikely to what was observed in the 2D case. At the latter stages, the rolls size decreases until their complete disappearance. The larger non-modal growth of 3D disturbances that reaches maximum at a later time, as compared to the 2D case, appears to be observed also for other shear flows (see, e.g., Schmid & Hennigson, 2001). A possible explanation of this seemingly common phenomenon is offered in Vitoshkin *et al.* (2012).

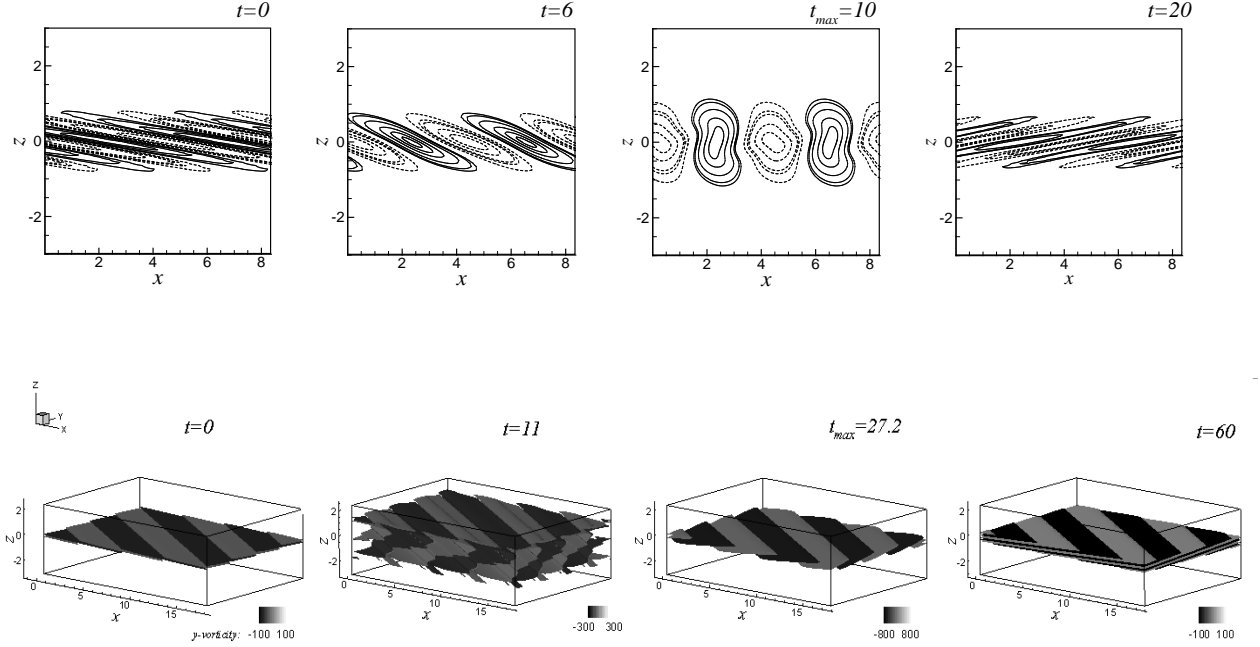

Fig. 8. Linear evolution of the spanwise vorticity component, $\eta_y$, of (a) two-dimensional and (b) three-dimensional optimal vectors. Upper frames: 2D case: $\alpha$=1.5, $\beta$=0, $Re$=1000. Lower frames: 3D case: $\alpha$ =0.7, $\beta$ =0.8, $Re$=1000.

*5.3. Non-linear effect on evolution of the optimal vector.*

To gather a better insight into time evolution of optimal disturbances, we consider also their fully non-linear development in time. Apparently, if the amplitude of optimal initial vector is small enough, the non-linear terms remain negligibly small during the whole time integration, so that the calculated flow resembles the predicted linear behavior. The growth of the initial perturbation kinetic energy coincides with one yielded by the IVP ODEs solution, as well as the one given by the growth function calculated by the SVD-based approach. This observation completes our verification of the non-modal growth results (see Fig.3).

To visualize the mixing layer flow, we follow Roger & Moser (1992) and Kit *et al.* (2010), and add a passively advected dimensionless temperature *T* to our model. Initially, the temperature is the same as the velocity *tanh* profile. Fig. 10 illustrates development of the 2D mixing layer flow starting from an optimally perturbed base flow at the parameters corresponding to the linearly stable case, $Re$=1000, $\alpha$=1.5, $\beta$=0. The two cases presented in Fig. 9 correspond to two different amplitudes of the optimal disturbance vector. The optimal vector, whose kinetic energy norm is unity, appears to be small enough, compared to the base flow, to exhibit fully linear behavior over the whole time interval considered. It is shown by the solid line in Fig. 9. Increasing the initial amplitude by the factor of 10 (dash line in Fig. 9) we observe linear behavior at early times $t<4$. At later times the increased amplitude makes the non-linear term non-negligible, which leads to a qualitatively different flow evolution that includes also a considerably smaller growth of kinetic energy, which, conversely to the linear predictions, attains several minima and maxima.

To illustrate qualitatively different flow patterns, we compare the snapshots of the passively advected temperature and the spanwise vorticity component in Figs. 10 and 11, respectively. The evolution of $T$ in the fully linear case (Fig. 10a) exhibits initial growth of zigzag wavy structures that disappear at longer times. The area where the passive scalar is completely mixed corresponds to the green color. An interesting observation is a wider zone of mixed $T$ at large times compared to that in the initial state. A relatively fast mixing is achieved here together with the complete disappearance of the initial perturbation and laminarization of the perturbed flow.

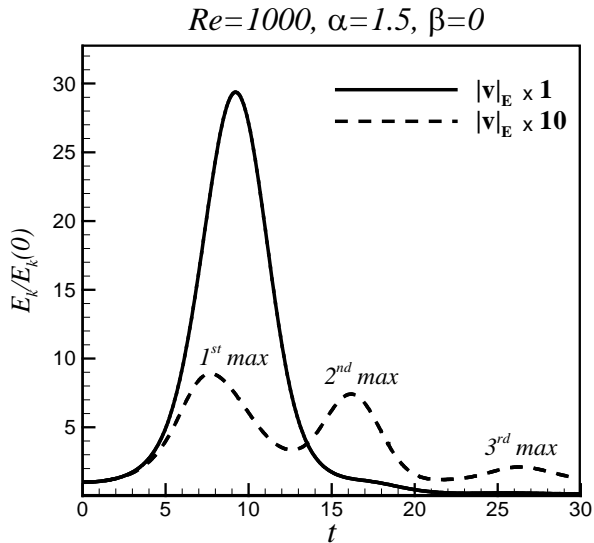

Fig. 9. Kinetic energy growth of the optimal vector with (dashed line) and without (solid line) non-linear effect

When the evolution of the non-modal disturbance switch on the non-linear terms (Fig. 9b) one observes appearance of the mushroom-shape structures at $t=7$ and 10. These structures look similar to the patterns of concentration reported in Figs. 13 and 14 of Lin & Corcos (1984) for a non-linearly developing mixing layer. Similar non-regular structures were also observed in the streamwise plane experimentally by Bernall & Roshko (1986). At a later time, $t=14$, the pattern becomes similar to the one reported by Smyth & Peltier (1991) in their Fig. 8. At a later time, $t=17$, we observe the patterns with elongated borders, which resemble classical experimental observations of Thorpe (1971) at long times (see his Figs. 8 and 9). Obviously, after the optimal disturbance decays, the passive scalar appears to be mixed even stronger than in the previous case, while the flow attains the plane-parallel velocity profile.

(a) Linear evolution

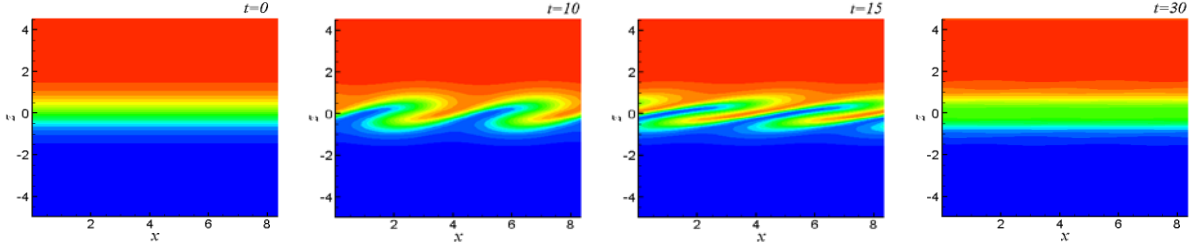

(b) Non-linear effect

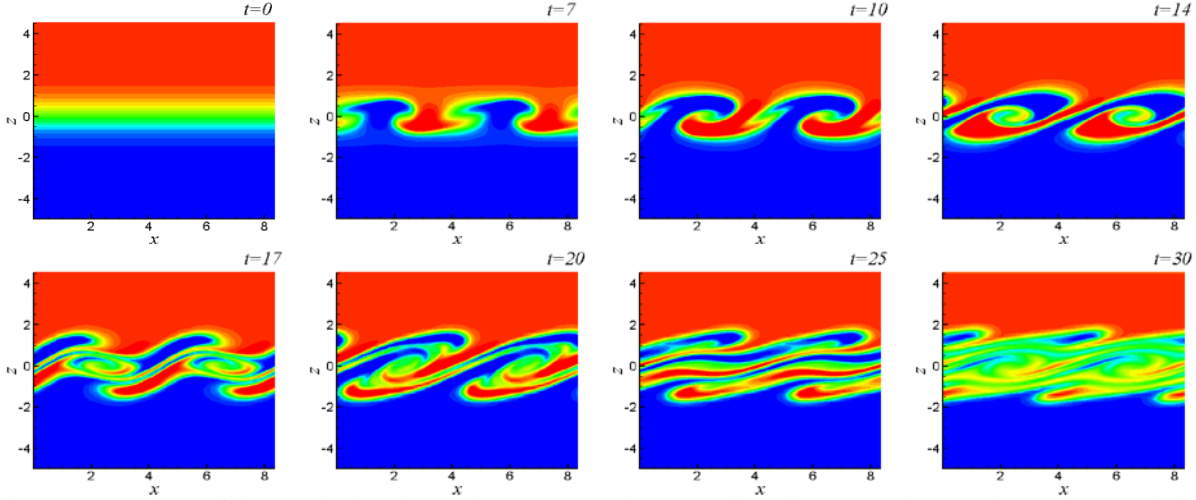

Fig. 10. Snapshots of the passively advected temperature in cases of (a) linear and (b) non-linear time evolution of the optimal disturbance. $Re=1000$, $\alpha=1.5$, $\beta=0$. Initial amplitude in the case (b) is 10 times larger than that of the case (a).

Snapshots of the spanwise vorticity are shown in Fig. 11. At early times the vorticity pattern consists of structures tilted against the shear slope, which we attribute to the Orr mechanism. At a later, time we observe that vorticity patterns are rotated, apparently by the base flow, around their centers located at the midplane. Again, we see some similarities of the vorticity behavior with the fully non-linear results reported, e.g., by Rogers & Moser (1992) and Smyth & Peltier (1991).

It should be noted that we are examining transient growth of optimal disturbance in order to discover transition of a parametrically stable flow to unstable regime. To do this we increase the amplitude of the initial optimal perturbation so non-linear terms will be triggered at later stages of the time evolution. The present calculations show that these non-linear effects do not lead to a noticeable sub-unstable transition. However, the optimal perturbation evaluates by different way than a non-linear evolution of perturbation based on the most unstable linear eigenmode, or KH mode, in unstable flow regime. As a calculation test, we performed non-linear computations using KH mode as initial vector and observed well-known evolution of the spanwise vortices described, e.g., in Ho & Huerre (1984). As it is seen from Figure 4.2.14, the evolution of the optimal vector is governed mainly by Orr-mechanism and differs qualitatively from the evolution of the leading linearly unstable mode.

Comparing the vorticity perturbation pattern with the change of the growth function (cf. Figs. 9 and 11), we observe that when the kinetic energy norm reaches the minimum, the vorticity isolines are elongated along the shear slope. At later times they turn against the shear, which leads to the next temporal growth. The maximal values of the kinetic energy norm correspond to the vorticity patterns elongated vertically, exactly as it was observed for the linearized problem.

Similarities between the computed flow structures and those observed in previous experimental and numerical studies allow us to make the following assumption. At late stages of time-development, the actual width of the mixing layer grows, thus leading to the growth of the dimensionless streamwise wavenumber $\alpha$. This necessarily results in a stabilization of the mixing layer flow. However, at this stage the flow is already strongly perturbed. Therefore, it is possible that development of the mixing layer flow at late stages is governed or strongly affected by the non-modal growth. The latter results in the flow structures similar to the ones observed here.

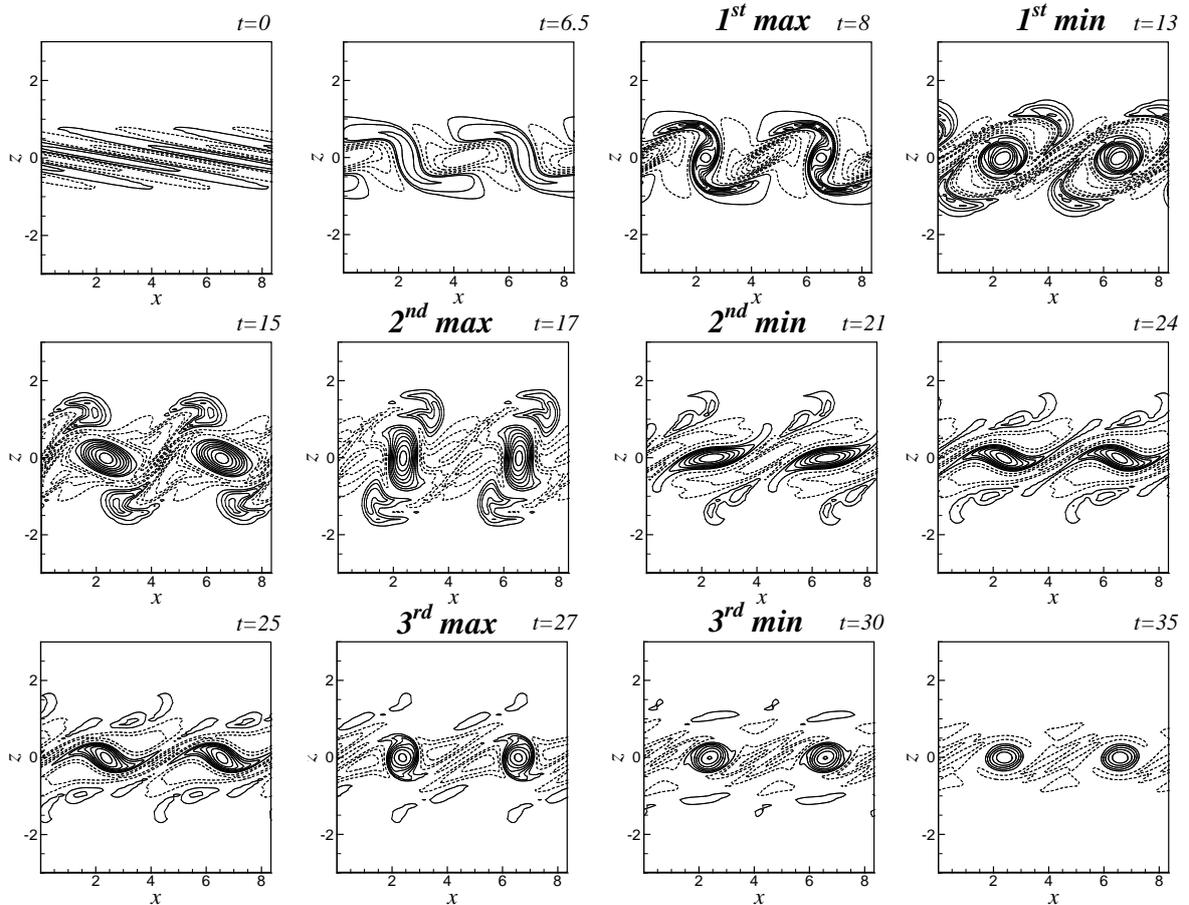

Fig. 11. Snapshots of the spanwise vorticity during non-linear time evolution of the optimal disturbance. $Re=1000$, $\alpha=1.5$, $\beta=0$.

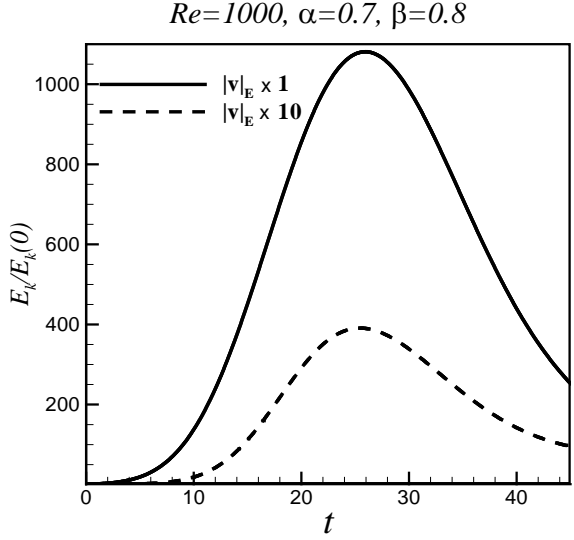

Fig. 12. Kinetic energy growth of a 3D optimal vector with and without non-linear effect

Time dependence of the kinetic energy norm for the 3D optimal disturbance is shown in Fig. 12. Similarly to the 2D case, the disturbance having the unity kinetic energy norm exhibits a completely linear behavior, and non-linear mechanisms switch on when the amplitude is increased by the factor of 10. However, in the 3D case the maximal growth of kinetic energy is attained at a considerably longer time (cf. Figs. 5 and 12), as is predicted by the above non-modal analysis. Note that in the 3D case, we do not observe several maxima and minima in the kinetic energy time history. Time evolution of the passively advected temperature is shown in Fig. 13. It generally resembles the structures observed in the 2D case, however the whole pattern is aligned along the disturbance vector $(2\pi/\alpha, 2\pi/\beta)$. After the perturbation decays, the width of mixed temperature zone is even larger than that observed in the 2D case. The qualitative difference of the 2D and 3D non-modal growth can be seen by comparison of Figs. 11 and 14. Figure 14 shows the snapshots of spanwise vorticity in the spanwise midplane. Similarly to the 2D case, the 3D growth starts from the optimal perturbation aligned against the shear slope. However, when the perturbation becomes aligned along the shear ($t \geq 20$ in Fig. 14), the kinetic energy continues to grow, thus leading to a larger growth at a longer time. This phenomenon seems to be common for all plane parallel shear flows and is addressed in a companion paper by Vitoshkin *et al.* (2013).

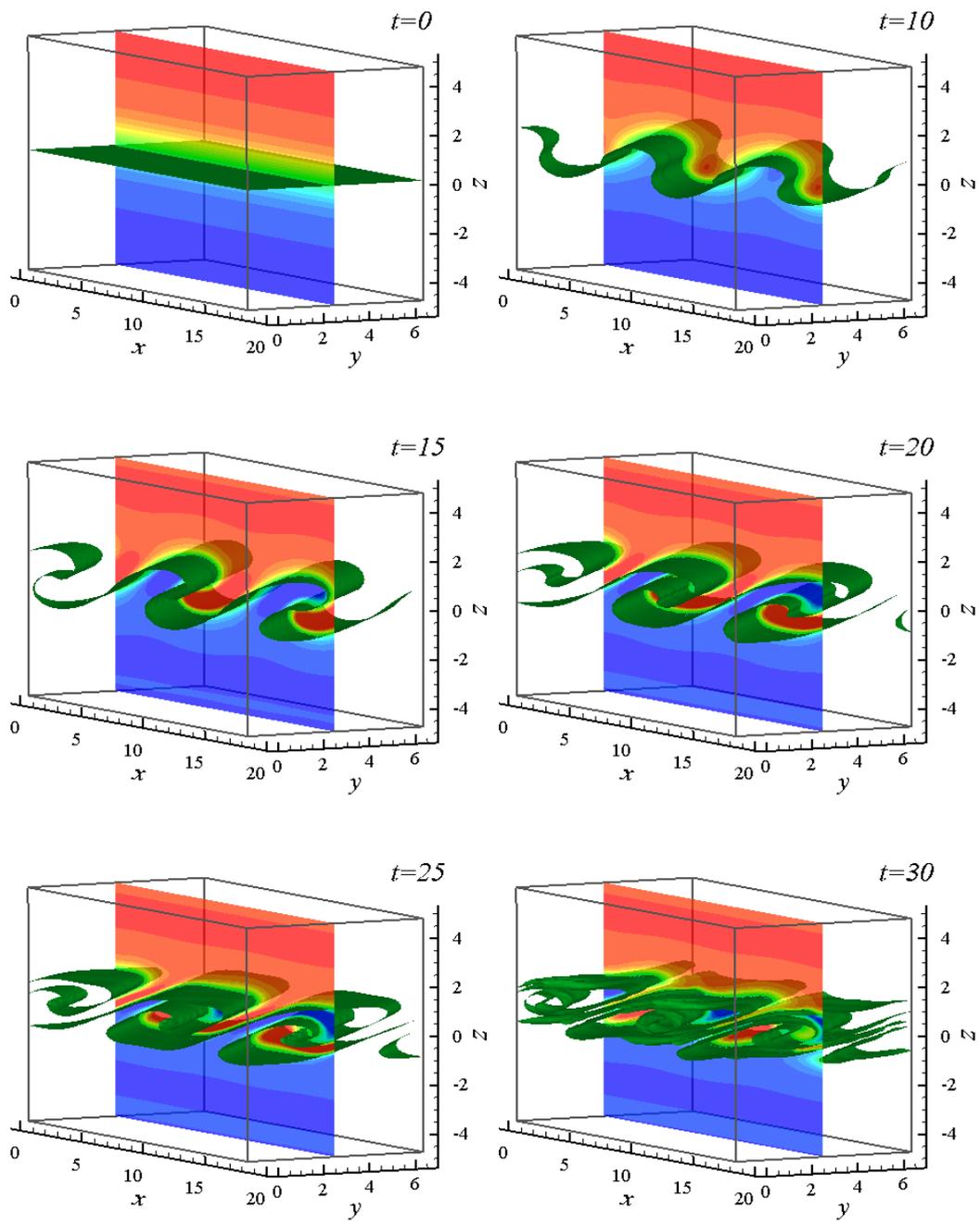

Fig. 13. Snapshots of the passively advected temperature during non-linear time evolution of a 3D optimal disturbance. $Re$=1000, $\alpha$=0.7, $\beta$=0.8.

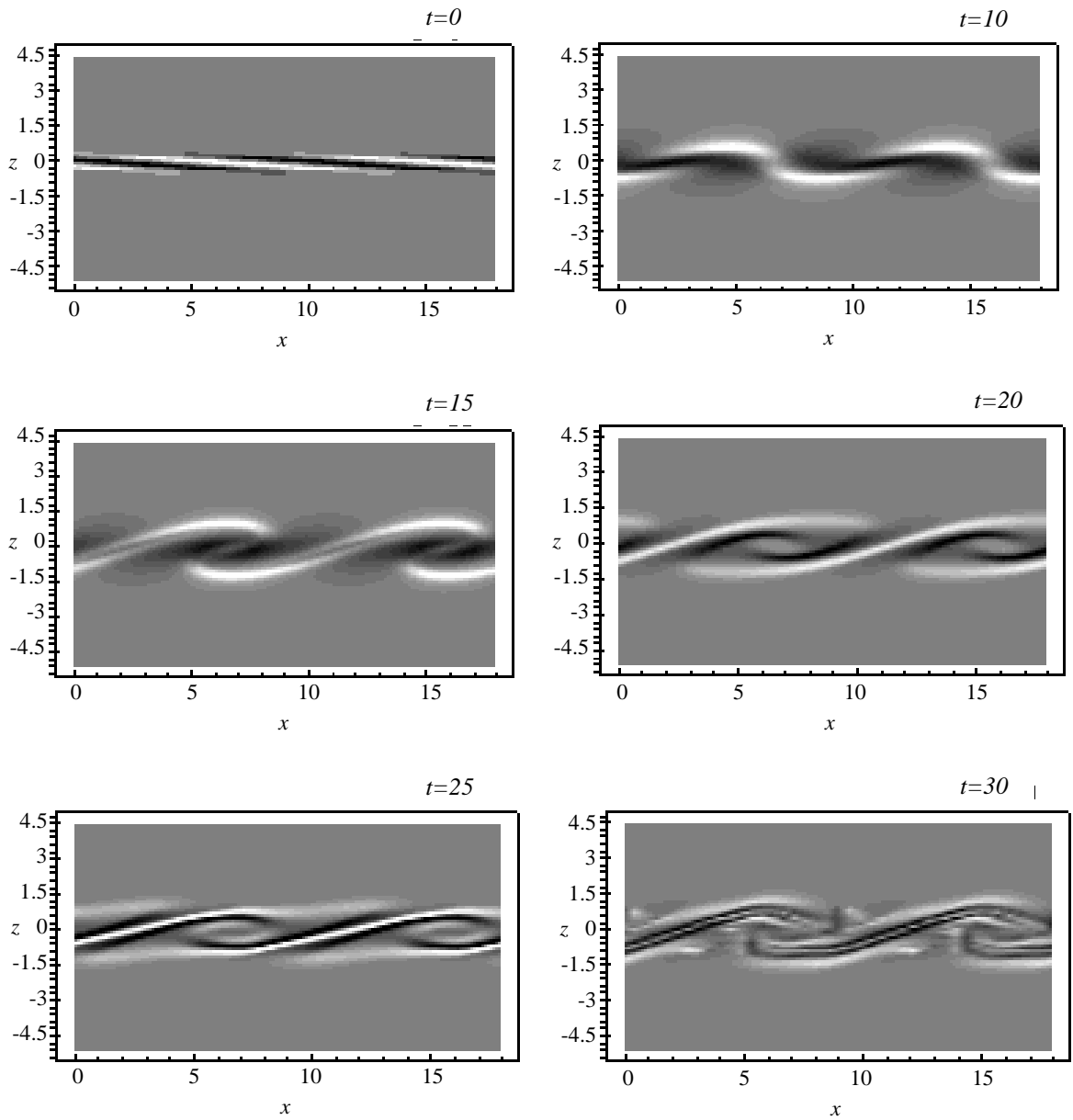

Fig. 14. Snapshots of the spanwise vorticity in the spanwise midplane during non-linear time evolution of the optimal disturbance. $Re=1000$, $\alpha=0.7$, $\beta=0.8$.

## 6. Conclusions

In the current study the transient perturbation dynamics in isothermal viscous mixing layers was investigated. The research combined fundamental theoretical principles along with the computational modelling.

The flow of interest exhibits significant transient growth, typically with many orders of magnitude of streamwise and spanwise wavenumbers, for which the flow is asymptotically stable. Comparing the calculated flow structures with those observed in several previous experimental and numerical studies, we speculate that the mixing layer flow at late stages of the linear instability development can be strongly affected by the non-modal disturbances growth. The optimal perturbation is always localized inside the shear zone.

The following conclusions were drawn from the investigation of transient growth in an isothermal viscous mixing layer:

- A series of numerical tests performed revealed that the mixing layer flow appears to be a more numerically challenging problem for the non-modal growth analysis than the problems considering bounded (e.g., plane Couette and Poiseuille flows) or semi-inbounded (e.g., Blasius boundary layer profile) flows. The correct numerical modeling of the non-modal growth for the $tanh$-velocity profile requires much better resolution in the cross-flow direction than other plane-parallel flows.
- A problem that may need attention is the separation of the discrete and continuous parts of the spectrum. We provided several arguments on why and how the discrete modes can be extracted and why only a discrete part of the spectrum should be taken into account when non-modal growth of disturbances in the isothermal mixing layer flow is studied. The corresponding results are verified by using three independent approaches for calculation of the growth function, as well as by the time-dependent calculations applied to both the ODE system resulting from the discretized Orr-Sommerfeld and Squire equations, and by the fully 3D non-linear time-dependent flow model.
- There is a possibility to obtain a significant mixing without making the flow turbulent. This observation is based on the advection of passive temperature and is observed during both 2D and 3D, linear and non-linear evolution of an optimally disturbed mixing layer flow. Therefore, the growth of small perturbations due to non-modal instability consequently may provide an additional possibility for mixing.
- A three-dimensional direct numerical simulation of the mixing layer flow which starts from the optimally perturbed base flow was conducted to investigate non-linear evolution of optimal perturbation and possibility for a by-pass transition. Following time non-linear evolution of the optimal disturbances, we observe qualitatively different development in 2D and 3D cases. In the 2D case, the non-linear effects lead to the appearance of several maxima and minima in the time history of the kinetic energy. Evolution of the spanwise vorticity pattern reveals that the minima are observed when the iso-vorticity lines are tilted along the shear slope. The temporal growth starts when the isolines become tilted against the shear, and the maxima are reached when they are rotated by the base flow until they become vertically aligned. No several minima or maxima are observed in the non-linear development of the 3D optimal disturbances. The maximum of the kinetic energy in the 3D case is attained significantly later, compared to the 2D case, after the patterns had turned in the base flow direction.

**A**cknowledgement

This study was supported by BSF (Bi-National US-Israeli Foundation) grant No. 2004087. The authors wish to express their thankfulness to E. Kit and E. Heifez for long and fruitful discussions of these results.

**REFERENCES**

Åkervik E., Ehrenstein U., Gallaire F., and Henningson D.S. 2008 Global two-dimensional stability measures on the flat plate boundary-later flow. *Eur. J. Mech. B/Fluids*, **27**, 501-513.

Andersson P., Berggren M., and Henningson D.S. 1999 Optimal disturbances and bypass transition in boundary layers, *Phys. Fluids*, **11**, 134-150.

Bakas N.A. and Ioannou P.J. 2009 Modal and nonmodal growths of inviscid planar perturbations in shear flows with a free surface. *Phys. Fluids*, **21**, 024102.

Bernal L.P. and Roshko A. 1986 Streamwise vortex structure in plane mixing layers. *J. Fluid Mech.*, **170**, 499-525.

Bun Y. and Criminale W.O. 1994 Early-period dynamics of an incompressible mixing layer. *J. Fluid Mech.*, **273**, 21-82.

Butler R.M. and Farrell B.F. 1992 Three-dimensional optimal perturbations in viscous shear flows, *Phys. Fluids A*, **4**, 1367-1650.

Corbett P. and Bottaro A. 2000 Optimal perturbations for boundary layers subject to streamwise pressure gradient. *Phys. Fluids*, **12**, 120-130.

Criminale W.O., Jackson T.L. and Lasseigne D. 1995 Towards enhancing and delaying disturbances in free shear flows, *J. Fluid Mech.*, **294**, 283-300.

Criminale W.O., Jakson T.L. and Joslin R.D. 2003 Theory and computation in Hydrodynamic Stability. Cambridge University Press.

Farrell B.F. 1987 Developing disturbances in shear. *J. Atmospheric Sci.*, **44**, 2191-2199.

Farrell B.F. 1988. Optimal excitation of perturbations in viscous shear flow. *Phys. Fluids*, **31**, 2093-2102.

Gaster M., Kit E., and Wygnanski I. 1985 Large-Scale Structures in a Forced Turbulent Mixing Layer. *J. Fluid Mech.*, **150**, 23-39.

Gelfgat A.Yu. and Kit E. 2006 Spatial versus temporal instabilities in a parametrically forced stratified mixing layer, *J. Fluid Mech.*, **552**, 189-227.

Grosch C.E. and Salwen H. 1978 The continuous spectrum of the Orr-Sommerfeld equation. Part 1. The spectrum and the eigenfunctions. *J. Fluid Mech.*, **87**, 33-54.

Hazel P. 1972 Numerical studies of the stability of inviscid stratified shear flows. . *J. Fluid Mech.*, **51**, 39-61.

Healey J. J. 2009 Destabilizing effects of confinement on homogeneous mixing layers. *J. Fluid Mech.*, **623**, 241–271.

Heifetz E. and Methven J. 2005 Relating optimal growth to counterpropagating Rossby wavesin shear instability. *Phys. Fluids*, **17**, 064107.

Ho C.-M., Huerre P. Perturbed free shear layers. *Ann. Rev. Fluid Mech.*, **16**, 365–424, 1984.

Joseph D.D. 1976. Stability of Fluid Motions. Springer-Verlag, New York.

Kit E., Wygnansky I., Friedman D., Krivonosova O., and Zhilenko D. 2007 On the periodically excited plane turbulent mixing layer, emanating from a jagged partition. *J. Fluid Mech.*, **589**, 479-507.

Kit E., Gerstenfeld D., Gelfgat A. Y., and Nikitin N.V. 2010 Bulging and bending of Kelvin-Helmholtz billows controlled by symmetry and phase of initial perturbation. *J. Physics: Conference Series,* **216**, 012019.

Le Dizès S. 2003 Modal growth and non-modal growth in a stretched shear layer. *Eur. J. Mech.*, B/Fluids, **22**, 411-430.

Lin S.J. and Corcos G.M. 1984 The mixing layer: deterministic models of a turbulent flow. Part 3. The effect of plane strain on the dynamics of streamwise vortices. *J. Fluid Mech.*, **141**, 139-178.

Mao X. and Sherwin S. J. 2012 Transient growth associated with continuous spectra of the Batchelor vortex. *J. Fluid Mech.*, **697,** 35-59.

Reddy S.C. and Henningson D.S. 1993 Energy growth in viscous channel flows, *J. Fluid Mech.*, **252**, 209-238.

Rogers M. M. and Moser R. D. 1992 The three-dimensional evolution of a plane mixing layer: the Kelvin-Helmholtz rollup. *J. Fluid Mech.*, **243,** 183-226.

Schmid P.J. 2000 Linear stability theory and bypass transition in shear flows. *Physics of Plasmas*, **7**, 1788-1794.

Schmid P.J. and Henningson D.S. *Stability and transition in shear flows*. Springer, N.Y., 2001.

Smyth W.D. and Peltier W.R. 1991 Instability and transition in finite-amplitude Kelvin-Helmhotz and Holmboe waves. *J. Fluid Mech.*, **228**, 387-415.

Thorpe S.A. 1971 Experiments on the instability of stratified shear flows: miscible fluids. *J. Fluid Mech.*, **46**, 299-319.

Trefethen L. N. and Embree M. 2005 Spectra and Pseudospectra: The Behavior of Nonnormal Matrices and Operators**.** Princeton University Press, 624 pp.

Vitoshkin H., and Gelfgat A. Y. 2013 On direct inverse of Stokes, Helmholtz and Laplacian operators in view of time-stepper-based Newton and Arnoldi solvers in incompressible CFD. *Submitted for publication*. arXiv:1107.2461v1

Vitoshkin H., Heifetz E., Gelfgat A. Y., Harnik N. 2013 On the role of vorticity stretching in optimal growth of three dimensional perturbations on plane parallel shear flows. *J. Fluid Mech.*, **707**, 369-380.

Yecko P., Zaleski S., and Fullana J.-M. 2002 Viscous modes in two-phase mixing layers. *Phys. Fluids*, **14**, 4115-4123.

Yecko P. and Zaleski S. 2005 Transient growth in two-phase mixing layers. J. *Fluid Mech.*, **528**, 43-52.

# APPENDIX A – Estimation of the effect of viscous dissipation effect on the plane-parallel mixing layer flow

The linearized Orr-Sommerfeld and Squire equations assume that the plane-parallel mixing layer flow is frozen. This is correct only within the inviscid flow model. In the viscous case the *tanh*-mixing layer width is continuously growing, which is most profound at small Reynolds numbers, i.e., at large viscosities. In the following, we present a simple model allowing us to estimate until which times the viscosity effect can be neglected.

Consider the unsteady momentum equation with initial velocity **u**(0,0,*u(t,z)*), where *u(t=0,z)* has the hyperbolic tangent profile. We want to estimate how the mixing layer thickness $\delta_v$ varies, owing to viscosity, with time at different Reynolds numbers. Assuming that the flow remains pane-parallel, we arrive at a problem similar to the one dimensional heat conduction equation:

$$\frac{\partial u}{\partial t} = \frac{1}{Re}\frac{\partial^2 u}{\partial z^2}, \tag{A.1}$$

$$u(t=0) = tanh(z), u(L) = 1, u(-L) = -1, \tag{A.2}$$

The solution of (A.1)-(A.2) is obtained by the standard separation of variables

$$u(t,z) = 1 + \frac{z+L}{L} + \frac{2}{\pi}\sum_1^\infty \frac{\cos(n\pi)+1}{n}\sin\frac{n\pi(z+L)}{2L} e^{-\frac{n^2\pi^2 t}{4L^2 Re}}$$

$$+ \frac{1}{L}\sum_1^\infty \int_0^{2L}\left(\tanh(\xi - L)\sin\frac{n\pi\xi}{2L}\right)d\xi \sin\frac{n\pi(z+L)}{2L} e^{-\frac{n^2\pi^2 t}{4L^2 Re}}, \tag{A.3}$$

The shear layer thickness is defined as a distance between the points at which the velocity attains the values of ±0.99. The result is presented in Figure A1. Obviously, for *Re*≤10 the layer thickness strongly depends on the Reynolds number. For 10≤*Re*≤100 the linearized problem is meaningful only at very short times, *t*<2, however, it already allows considering the linear stability of the flow. For *Re*>100 the thickness remains almost unchanged until *t*=30. Note that the reported maximal values of the non-modal growth functions were reached within this time interval.

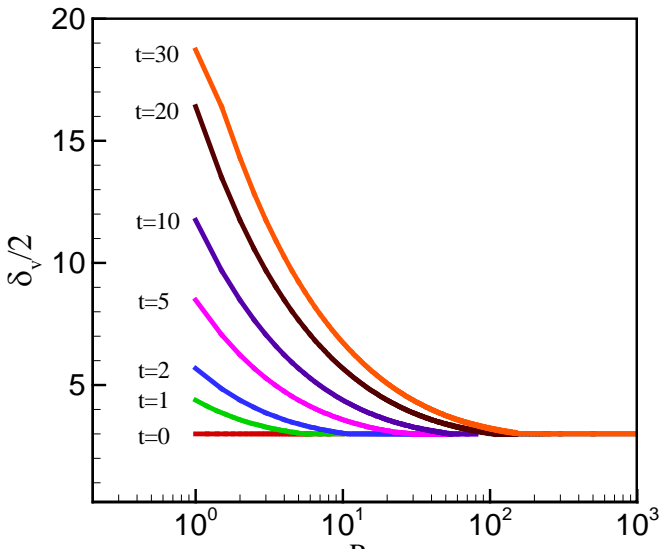

Fig. A1. Values of the shear layer thickness as a function of the Reynolds number and time.

# APPENDIX B – Eigenproblem and complex conjugated eigenvalues

Consider the eigenproblem related to equations (2) and (3). Assuming $w(z,t) = w(z)e^{\lambda t}$, $\eta(z,t) = \eta(z)e^{\lambda t}$ we obtain

$$\lambda(w'' - k^2 w) = i\alpha[U''w - U(w'' - k^2 w)] - \frac{1}{Re}(w'''' - 2k^2 w'' + k^4 w), \quad (B.1)$$

$$\lambda\eta = -i\beta U'w + \frac{1}{Re}(\eta'' - k^2\eta) - i\alpha U\eta, \quad (B.2)$$

where the tag denotes the derivative with respect to $z$. Assume also that $\lambda = \lambda_r + i\lambda_i$ is an eigenvalue, and $w(z) = w_r(z) + iw_i(z)$, $\eta(z) = \eta_r(z) + i\eta_i(z)$ are the corresponding eigenfunctions of (B.1, B.2). Substituting these expressions back into equations (B.1), (B.2) and separating real and imaginary parts yields the two following pairs of equations:

$$\lambda_r(w_r'' - k^2 w_r) - \lambda_i(w_i'' - k^2 w_i) =$$
$$= -\alpha[U''w_i - U(w_i'' - k^2 w_i)] - \frac{1}{Re}(w_r'''' - 2k^2 w_r'' + k^4 w_r) \quad (B.3)$$

$$\lambda_i(w_r'' - k^2 w_r) + \lambda_r(w_i'' - k^2 w_i) =$$
$$= \alpha[U''w_r - U(w_r'' - k^2 w_r)] - \frac{1}{Re}(w_i'''' - 2k^2 w_i'' + k^4 w_i) \quad (B.4)$$

$$\lambda_r \eta_r - \lambda_i \eta_i = \beta U' w_i + \frac{1}{Re}(\eta_r'' - k^2 \eta_r) + \alpha U \eta_i \quad (B.5)$$

$$\lambda_r \eta_i + \lambda_i \eta_r = -\beta U' w_r + \frac{1}{Re}(\eta_i'' - k^2 \eta_i) - \alpha U \eta_r \quad (B.6)$$

We notice that due to the fact that *tanh(z)* is an odd function of *z*, the transformation $z \to -z$, $U(z) = -U(-z)$ results in the same problem. Applying this transformation to equations (B.3)-(B.6) yields

$$\hat{\lambda}_r(\hat{w}_r'' - k^2 \hat{w}_r) - \hat{\lambda}_i(\hat{w}_i'' - k^2 \hat{w}_i) =$$
$$= \alpha[U''\hat{w}_i - U(\hat{w}_i'' - k^2 \hat{w}_i)] - \frac{1}{Re}(\hat{w}_r'''' - 2k^2 \hat{w}_r'' + k^4 \hat{w}_r) \quad (B.7)$$

$$\hat{\lambda}_i(\hat{w}_r'' - k^2 \hat{w}_r) + \hat{\lambda}_r(\hat{w}_i'' - k^2 \hat{w}_i) =$$
$$= -\alpha[U''\hat{w}_r - U(\hat{w}_r'' - k^2 \hat{w}_r)] - \frac{1}{Re}(\hat{w}_i'''' - 2k^2 \hat{w}_i'' + k^4 \hat{w}_i) \quad (B.8)$$

$$\hat{\lambda}_r \hat{\eta}_r - \hat{\lambda}_i \hat{\eta}_i = -\beta U' \hat{w}_i + \frac{1}{Re}(\hat{\eta}_r'' - k^2 \hat{\eta}_r) - \alpha U \hat{\eta}_i \quad (B.9)$$

$$\hat{\lambda}_r \hat{\eta}_i + \hat{\lambda}_i \hat{\eta}_r = \beta U' \hat{w}_r + \frac{1}{Re}(\hat{\eta}_i'' - k^2 \hat{\eta}_i) + \alpha U \hat{\eta}_r \quad (B.10)$$

where all the functions depend on $(-z)$. The problems (B.3)-(B.6) and (B.7)-(B.10) are identical and have the same solutions. It is easy to see that choosing $\hat{\lambda}_r = \lambda_r$, $\hat{\lambda}_i = -\lambda_i$, $\hat{w}_r(-z) = w_r(z)$, $\hat{w}_i(-z) = -w_i(z)$, $\hat{\eta}_r(-z) = \eta_r(z)$, $\hat{\eta}_i(-z) = -\eta_i(z)$, we arrive to the equations (B.3)-(B.6). Thus, if $\lambda = \lambda_r + i\lambda_i$ is the eigenvalue of (B.1), (B.2), then $\lambda = \lambda_r - i\lambda_i$ is also the eigenvalue. The corresponding eigenvectors are connected via reflection and antireflection of their real and imaginary parts, respectively, with respect to the plane *z=0*.

# APPENDIX C – Critical energetic Reynolds number for mixing layer flow.

To examine at which parameters the non-modal growth is possible we compute the energetic critical Reynolds number. The energetic critical Reynolds number $Re_E$ is defined as a minimal value of $Re$ at which there exists a possibility for the kinetic energy growth at the initial time, i.e., $dE/dt>0$ at $t=0$. Since we address a possibility of the energy growth at initial time only, the viscous dissipation effects described in the Appendix A are irrelevant. Following Reddy & Henningson (1993), we calculate $Re_E$ using the calculus of variations. The energetic critical Reynolds number $Re_E(\alpha,\beta)$ is equal to the smallest positive eigenvalue $\lambda$, of the coupled eigenvalue problem:

$$\frac{1}{\lambda}\begin{bmatrix}\Delta^2 & 0 \\ 0 & -\Delta\end{bmatrix}\begin{pmatrix}w \\ \eta\end{pmatrix} = \begin{bmatrix}-i\left(\alpha U'\frac{\partial}{\partial z} + 0.5\alpha U''\right) & i(0.5\beta U') \\ i0.5\beta U' & 0\end{bmatrix}\begin{pmatrix}w \\ \eta\end{pmatrix} \quad (C.1)$$

$$z = \pm L: w = w' = \eta = 0$$

Note that it follows from the formulation (C.1) that outside the mixing zone, where $U'=0$, both variables $\eta$ and $w$ vanish, which means that the continuous spectrum is excluded from the consideration also here.

Figure C1 shows level curves of the critical $Re_E$ in the $(\alpha, \beta)$ plane. The minimal critical Reynolds number for energy growth equals $\approx 0.51$, is reached at $\alpha=0$ and $\beta=0.09$, and increases with the increase of either $\alpha$ or $\beta$. Note, that also for Couette and Poiseuille flows the minimal value of $Re_E$ corresponds to $\alpha=0$ (Reddy & Henningson,1993), i.e. to a two-dimensional perturbation located in the spanwise plane. Similar observations are reported also by Yecko et al (2002) and Yecko & Zaleski (2005) for the two-phase mixing layer flow. The minimal critical energetic Reynolds number corresponding to the 2D disturbances is approximately 0.9. We conclude that the non-modal growth of kinetic energy of a perturbation is possible for $Re>0.51$, which is much smaller than usually considered cases and significantly smaller than the linear stability limit (Gelfgat & Kit, 2006). The graph shows that non-modal growth is possible also for $\alpha >1$, at which the mixing layer flow is stable for any Reynolds number and also for the inviscid case. The latter is consistent with the results reported. The energetic and linear critical Reynolds numbers corresponding to 2D perturbations are compared in Fig. C2a. The linear stability results are taken from Gelfgat & Kit (2006) and are rescaled according to the present formulation. Apparently, the $Re_E(\alpha,0)$ values are smaller than critical Reynolds numbers of the linear stability analysis. The energetic critical Reynolds numbers corresponding to different values of $\beta$ (3D initial perturbation) are shown in Fig. C2b. At small, but non-zero, values of $\beta$ the three-dimensional perturbations exhibit earlier initial growth than the two-dimensional ones corresponding to $\beta=0$. With the increase of $\beta$ the curves corresponding to different $\beta$ tend to coincide. This means that at the same value of $\alpha$ the initial growth can start as a superposition of several initial perturbations having different spanwise periodicity. This can be important for understanding experimental results, as well as for choice of the computational domain and initial states for computational simulations.

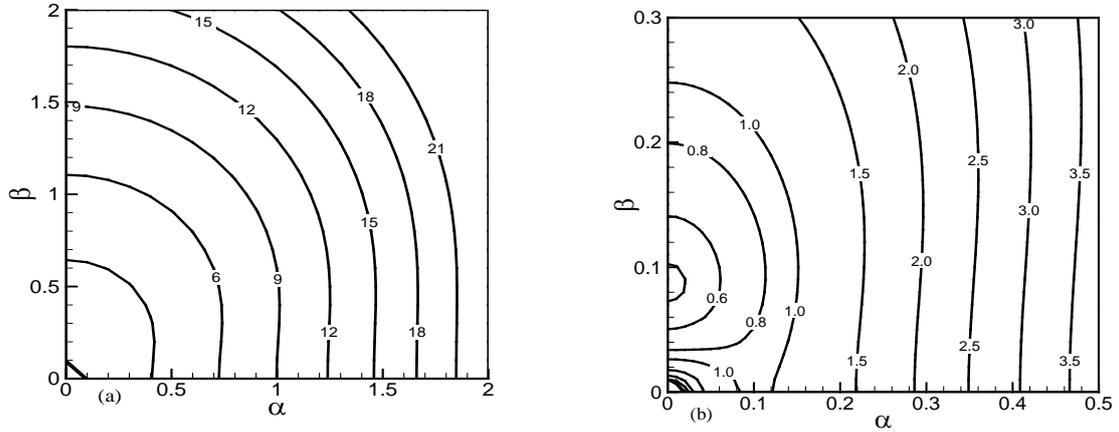

Fig. C1. Level curves of $Re_E(\alpha,\beta)$ for mixing layer flow. Frame (b) zooms in the lower left corner of frame (a).

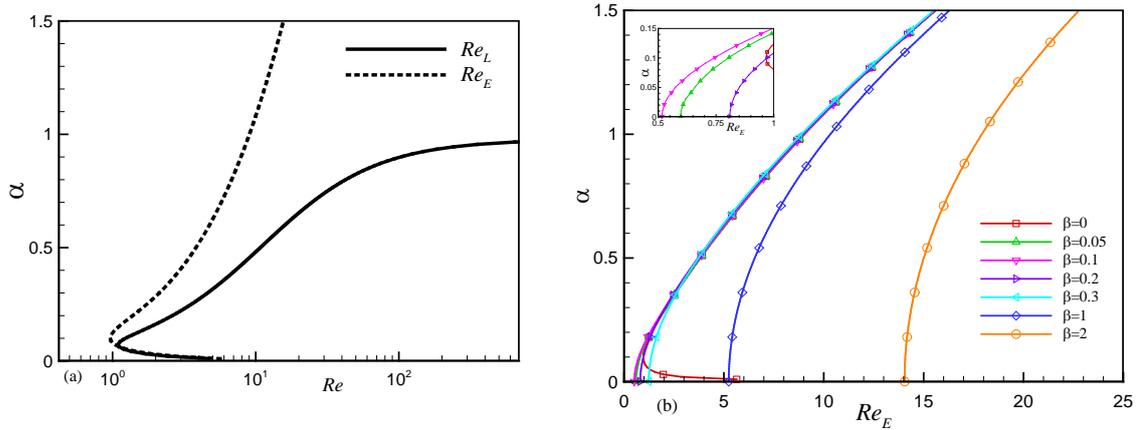

Fig. C2. (a) Critical linear and an energetic stability curves for two-dimensional disturbances ($\beta=0$). (b) Critical energetic Reynolds numbers $Re_E$ as functions of $\alpha$ and $\beta$.